\documentclass[11pt]{article}
\usepackage[margin=1.1in]{geometry}
\usepackage{amsmath,amssymb,amsthm}

\usepackage[colorlinks=true,linkcolor=blue,citecolor=blue,urlcolor=blue]{hyperref}

\newtheorem{theorem}{Theorem}
\newtheorem{lemma}{Lemma}
\newtheorem{conjecture}{Conjecture}
\newtheorem{proposition}{Proposition}
\theoremstyle{remark}

\newcommand{\I}{\mathbb{I}}
\newcommand{\ad}{\operatorname{ad}}
\newcommand{\im}{\operatorname{Im}}
\newcommand{\Q}{\mathcal{Q}}
\newcommand{\sq}{\sqrt{(N-25)(N-1)}}

\title{Commuting integrals of motion for free bosonic CFTs revisited:\\
certified densities, spectral rigidity, and an exact\\
identification at the $c=1$ branch point}
\author{Evgueniy Vitchev\\[3pt]
{\small Independent researcher, Greater Philadelphia,
Pennsylvania, USA}\\
{\small \texttt{evgueniy.vitchev@rutgers.edu} \;\;
ORCID: \href{https://orcid.org/0009-0000-2613-454X}{0009-0000-2613-454X}}}
\date{August 2026}

\begin{document}
\maketitle

\begin{abstract}
We revisit the commuting local integrals of motion for
free bosonic CFTs constructed in \cite{Vitchev} as input data for
integrable boundary states, using exact-arithmetic operator-product
engines. A bootstrap lemma shows that the Jacobi identity reduces the
overdetermined bilinear commutativity systems of \cite{Vitchev} to a
choice of family-dependent distinguished odd charge $D$ plus linear
kernel problems --- conditionally on \emph{even-spin vacancy}, a
machine-verified but finite-range input. For the
$\mathbb{Z}_2\times\mathbb{Z}_2$ ``paperclip'' family we publish the
previously unavailable spin-7 density $P_8$ in closed form, prove
$[\I_3,\I_5]=[\I_3,\I_7]=[\I_5,\I_7]=0$ identically in the parameter
$n$, and verify the Lukyanov--Vitchev--Zamolodchikov four-screening
characterization \cite{LVZ} at spins 7 and 9 at sampled $n$. For the
$A_2^{(2)}$ family we give an exact Coulomb-gas dictionary to the dual
$\Phi_{1,2}/\Phi_{2,1}$ perturbations (that identification itself
being classical \cite{ZamCount,Smirnov}), whence the infinite
hierarchy at all spins $6n\pm1$ follows by the Feigin--Frenkel
affine-Toda theorem; our contribution is the certified dictionary, not
a new all-spin construction. At $c=1$ the tower's defining
charges are identified exactly with quantum KdV at
$c_{1,4}=-\tfrac{25}{2}$ --- the all-odd sector of the
Freeman--Hornfeck--West family \cite{FHW}: the hierarchy embeds in
the enhanced all-odd tower at every odd spin, with equality verified
through spin 15 and all-spin multiplicity one open.  We exclude or
disfavor the named non-KdV candidates at the tested loci and identify
the self-dual degeneracy as $L(1,0)$. Through spin 11, the generic centralizer
of $\I_3$ is, for all three $\mathbb{Z}_2\times O(N)$
``cylindrical'' families, one-dimensional at odd spins and zero at
even spins, with
$\mathbb{Q}(t)$-symbolic charges at spins 7, 9, 11 and an explained
fiber jump at the decoupling point $N=-2$. We also certify the
published low-spin formulas of \cite{Vitchev} and record five
corrections (one presumed). All-spin statements are made only where proved and are
left finite-range otherwise. We close with a corank--character
conjecture linking the rank deficiencies of the bilinear systems to
characters of screening kernels.
\end{abstract}

\section{Introduction and summary of results}

Families of mutually commuting local integrals of motion (IM) of a
bulk conformal field theory are the defining data of integrable
boundary conditions: a boundary state $|B\rangle$ compatible with the
family satisfies $(\I_s-\bar\I_s)|B\rangle=0$, and boundary operators
and partition functions then admit expansions governed by the IM
eigenvalues \cite{GZ,BLZ1}. For free bosonic theories, the
earlier paper \cite{Vitchev} developed a direct computational
approach: enumerate
candidate densities of fixed spin and symmetry, reduce modulo total
derivatives, and impose vanishing of operator-product residues,
producing \emph{bilinear} homogeneous systems for the coefficient
vectors. That search recovered the KdV--Virasoro series, the
``$O(N)$'' series of the circular-brane model \cite{LZcirc}, an
$A_2^{(2)}$ (``$\Phi_{1,2}$'') series, the
$\mathbb{Z}_2\times\mathbb{Z}_2$ family of the paperclip model
\cite{LVZ}, and three new $\mathbb{Z}_2\times O(N)$ ``cylindrical''
families. Two puzzles were left open in \cite{Vitchev}: the bilinear
systems are heavily overdetermined yet consistently solvable, with no
structural explanation; and the existence of the towers past the
computed spins was not established there.  For the $A_2^{(2)}$ family
the missing all-spin step can in fact be supplied by the earlier
Feigin--Frenkel affine-Toda theorem once the normalization dictionary
is made explicit (Section~\ref{sec:a2}).  For the circular/$O(N)$,
paperclip, and cylindrical families the problem remains open: screening
\emph{proposals} exist for the paperclip and pillow-brane families
\cite{LVZ,LZpillow} and for circular-brane-type and BCD systems
\cite{LitSpod,LitVil}, but proofs of infinitude are repeatedly and
explicitly declared missing (see \cite{LZpillow}), the rigorous
technology being available only in the Toda/affine setting through the
Feigin--Frenkel cohomological argument \cite{FF}; the
$W_3$/Boussinesq construction \cite{BHK} supplies an independent
$A_2^{(2)}$ transfer-matrix cross-check.

This note reports a computational-and-structural pass over all the
families of \cite{Vitchev} in exact arithmetic. Our results:

\begin{enumerate}
\item[\bf R1.] \emph{Solvability explained, conditionally on
even-spin vacancy.} An elementary bootstrap
lemma (Section~\ref{sec:lemma}): once a distinguished odd-spin charge
$D$ is fixed --- $\I_3$ for the KdV, paperclip and cylindrical
families, $\I_5$ for $A_2^{(2)}$ --- the higher charges solve the
\emph{linear} systems $[D,\I_\sigma]=0$, and if no even-spin charge
commutes with $D$ (\emph{even-spin vacancy}) the Jacobi identity
forces the remaining higher--higher commutators
$[\I_{\sigma_1},\I_{\sigma_2}]$ to vanish identically. The genuinely
nonlinear content of the bilinear systems is only the choice of $D$;
the systems $[D,\I_\sigma]=0$ are genuine linear constraints, and it
is the higher--higher bilinear equations that become redundant,
within the range where vacancy is established. We verify the vacancy and
multiplicity-one hypotheses in exact arithmetic through charge spin
$10$ for the KdV and paperclip branches (Table~\ref{tab:spectral}),
through spin $13$ for $A_2^{(2)}$, and through spin $11$ for the
cylindrical families.
\item[\bf R2.] \emph{The paperclip $P_8$.} The previously unpublished
spin-7 paperclip density, as explicit rational functions of $n$
(Appendix~\ref{app:P8}).  Normalizing the $(\partial X)^8$
coefficient to $1$, the extremal $(\partial Y)^8/(\partial X)^8$
transition coefficient has zeros and poles at the
six resonance factors $(7n+2j)$ of the quasiclassical coefficient
$S_7$ of the paperclip ODE \cite{Vitchev}: poles at $j=4,5,6$ (and an
$n^3$ pole at $n=0$), zeros at $j=1,2,3$ (and an $(n+2)^3$ zero at
$n=-2$, dual to the $n^3$ pole), the two triples exchanged by the duality
$n\to-2-n$, under which the charge class is covariant modulo total
derivatives and overall normalization.
\item[\bf R3.] \emph{Exact commutativity.}
$[\I_3,\I_5]=[\I_3,\I_7]=0$ identically in $n$ (symbolic
verification), and $[\I_5,\I_7]=0$ identically in $n$ by exact
evaluation at $20$ rational points combined with an a priori degree
bound: with denominators cleared, every entry of the reduced
commutator is a polynomial in $n$ of degree at most $5+9=14$, so
vanishing at more than $14$ points is an identity.
$\I_9$ is re-derived with an exact certificate at sampled $n$
(existence and uniqueness of $\I_7,\I_9$ were reported in \cite{LVZ};
the closed-form $P_8$ and the reproducible certificates are the new
content).
\item[\bf R4.] \emph{Screening verification at spins 7 and 9.} Fitting
vertex operators $V_\alpha=e^{\alpha\cdot X}$ against
$[\oint V_\alpha,\I_3]=[\oint V_\alpha,\I_5]=0$ alone returns
the four vectors $\alpha=(\pm\sqrt{-n/2},\pm\sqrt{(n+2)/2})$ --- the
known screenings of \cite{LVZ} in the normalization of \cite{Vitchev}
--- and all four annihilate the independently computed $\oint P_8$
and, at sampled $n$, the newly computed $\oint P_{10}$ (49-sample
rational reconstruction, multiplicity one at each sample; its
denominator extends the resonance-factor pattern of $P_8$ to
$(9n{+}2j)$, $j=5,\dots,8$; see the artifact for the closed form).
The screening vectors are recovered from the lower charges on the
generic pivot stratum; the annihilation of $\oint P_8$ and
$\oint P_{10}$ is verified exactly at the stated sample values of $n$,
not as an identity in $n$.  With that scope, this is an independent
exact-arithmetic verification at spins 7 and 9 of the
four-screening characterization (eq.~(84) of \cite{LVZ}, left unproven
there; LVZ report low-spin checks). We claim verification of the LVZ
vectors, not exhaustiveness: at the special point $n=2$ the same fit
returns two additional pure-$Y$ vectors $(0,\pm\sqrt2)$, and the
elimination works on the generic pivot stratum, so completeness would
require a branch-complete polynomial-system analysis.
\item[\bf R5.] \emph{All-spin $A_2^{(2)}$ theorem and the
branch-point mechanism.}  An exact background-charge screening
calculation identifies the two sheets of the $I_5$ of
\cite{Vitchev} with the
dual $\Phi_{1,2}/\Phi_{2,1}$ realizations of $A_2^{(2)}$.  The
Feigin--Frenkel affine-Toda theorem \cite{FF} then proves, at the
generic point of the double cover, an infinite mutually commuting
hierarchy with one generator at every spin $6n\pm1$.  Independently,
on $s^2=(N-25)(N-1)$, rationally parametrized by
$N=(t^2-25)/(t^2-1)$, $s=-24t/(t^2-1)$, our exact scan finds that
content through spin 13 --- the members at spins $11,13$ are new as
explicit kernel data. Vacancy at spins $3$ and $9$ is
generic by rank semicontinuity, and the spin-11 and spin-13 charges
are constructed as explicit $\mathbb{Q}(t)$-symbolic densities with
commutators verified identically on the parameter curve: generic
existence is proven, and with sampled dimension $\le1$ the generic
kernels are exactly one-dimensional. At $c=1$ ($N=1$) the family is known
(from the same computations at the component level) to enhance to a
tower with all-odd pattern through the verified range, containing a charge
$\I_3'=\oint[(\partial X)^4-\tfrac52\,\partial X\partial^3X]$
outside the $c=1$ KdV ray. We show
this enhancement occurs neither at the branch point $N=25$ (where the
formal invariant algebra is faithful) nor for the branch coefficients
$(N,s)=(1,0)$ in the formal algebra: within the verified range it is
driven by the
finite-$N$ relations of the one-boson invariant algebra.  Section~\ref{sec:c1} identifies the tower --- quantum KdV at
$c_{1,4}=-\tfrac{25}{2}$ in Feigin--Fuchs realization, the all-odd
quantum-KdV sector of the $S_4$ commuting family of
Freeman--Hornfeck--West \cite{FHW}; the new content is the
branch-point embedding --- and its
mechanism, a $\mathbb{Z}_2$ symmetrization of the twisted
screening set available exactly where the background charge
vanishes; since $c=25$ has $Q^2=-2$, the same mechanism explains
the certified absence of enhancement at the other branch point, in
agreement with the formal-algebra diagnosis. Fibers of the
commuting-tower family jump at \emph{degeneration loci} of the
invariant algebra; at $N=1$ the degeneration locus and the
Langlands-self-dual branch point coincide, both expressing the
vanishing of the background charge.
\item[\bf R6.] \emph{Cylindrical families diagnosed.} All three
$\mathbb{Z}_2\times O(N)$ families carry all-odd, multiplicity-one
generic centralizer spectra through spin 11 on the parameter curve
($\mathbb{Q}(t)$-symbolic densities at spins 7, 9, 11; even-spin
vacancy generic by semicontinuity): existence and
uniqueness of $\I_7,\I_9,\I_{11}$ is new. The exponent content
matches the $A_1^{(1)}/D^{(2)}$ pattern and is decisively not
$A_2^{(2)}$-type,
sharpening the search for their home among D-type boundary labels in
the framework of \cite{LitVil}. At the point $N=-2$, where the $X$--$Y$
couplings of Solutions 2 and 3 vanish simultaneously, the kernels
double: the densities decouple and the commutant factorizes --- a
further, fully explained instance of fiber jumping at a degeneration
locus.
\item[\bf R7.] \emph{Certification and errata.} The low-spin formulas
of \cite{Vitchev} were re-derived from scratch, with one stated
exception: the cylindrical $\I_5$ formulas (54), (56), (58) were not
encoded, so the fifth erratum is established for the $\I_3$ formulas
and only presumed for those (Appendix~\ref{app:errata}). Correct as printed:
eqs.~(28)--(29), (31)--(33) --- now certified identically in $N$ ---
(34), (36)--(37) with (43)--(45), (47)--(49). Corrections
(Appendix~\ref{app:errata}): eq.~(30) [$657\to675$], eq.~(35) [two
dropped minus signs], eq.~(50) [$P_6^{[i]}\to P_4^{[i]}$], eq.~(51)
[missing factor $n$], and a factor-of-two normalization of the
$T_Y^2$ composite implicit in eqs.~(53), (55), (57) (the encoded and
tested $I_3$ formulas).
\item[\bf R8.] \emph{Corank--character conjecture.} The kernel
dimensions in Table~\ref{tab:spectral} suggest that the rank
deficiency of the bilinear systems \emph{equals} the graded character
of the corresponding joint screening kernel
(Section~\ref{sec:corank}); combined with R4 this reduces infinitude
of the paperclip tower to a Feigin--Frenkel-type character computation
for its four-screening system.
\end{enumerate}

Because the evidence below is of several different kinds,
Table~\ref{tab:claims} records the status of each claim explicitly.

\begin{table}[htbp]\centering\footnotesize
\begin{tabular}{@{}lp{1.95in}p{2.9in}@{}}
\hline\hline
 & claim & status and scope actually established\\
\hline
R1 & bootstrap lemma: Jacobi reduces the bilinear systems to the
choice of $D$ plus linear kernel problems & \emph{theorem},
conditional on even-spin vacancy; vacancy and multiplicity one
verified in exact arithmetic through charge spin 10 (KdV, paperclip),
13 ($A_2^{(2)}$), 11 (cylindrical)\\
R2 & paperclip spin-7 density $P_8$ & \emph{closed form, identically
in $n$}; the quoted zeros and poles are those of the extremal
$(\partial Y)^8/(\partial X)^8$ transition coefficient\\
R3 & $[\I_3,\I_5]=[\I_3,\I_7]=0$ & \emph{identity in $n$}, symbolic\\
 & $[\I_5,\I_7]=0$ & \emph{identity in $n$}, by exact evaluation at
20 rational points against a degree bound\\
 & $\I_9$ & \emph{exact certificate at sampled $n$}; existence and
uniqueness already reported in \cite{LVZ}\\
R4 & four-screening characterization at spins 7 and 9 & \emph{exact
at sampled $n$}, not an identity in $n$; verification of the
\cite{LVZ} vectors, \emph{not} exhaustiveness --- the elimination
works on the generic pivot stratum\\
R5 & $A_2^{(2)}$: one generator at every spin $6n\pm1$ &
\emph{theorem, all spin}, at the generic point of the double cover,
via the certified dictionary and \cite{FF}\\
 & scan content through spin 13; the spin-11 and spin-13 members &
\emph{$\Q(t)$-symbolic}, with commutators verified identically on the
parameter curve\\
 & $c=1$: the tower's defining charges $=$ quantum KdV at
$c_{1,4}=-\tfrac{25}{2}$ & \emph{exact identities}
(Theorem~\ref{thm:ident}); the hierarchy embeds at every odd spin
(Theorem~\ref{thm:inf}), equality verified through $\sigma\le15$,
all-spin multiplicity one open\\
 & the enhanced tower has a charge at every odd spin (so the
enhancement itself is all-spin) &
\emph{theorem, all spin} (Theorem~\ref{thm:inf}); the
specialization step is Lemma~\ref{lem:prim}\\
 & no Getmanov embedding reproduces the tower at two-charge depth &
\emph{theorem} (Theorem~\ref{thm:getmanov}), \emph{axis-aligned
only}; tilted embeddings remain open\\
R6 & cylindrical: all-odd, multiplicity one through spin 11, with
$\I_7,\I_9,\I_{11}$ new & \emph{$\Q(t)$-symbolic}; even-spin vacancy
generic by semicontinuity; the cylindrical higher--higher
commutators carry GF($p$) certificates with tags $=$ $d$-rank
(18 of the inventory's 21 instances; the other three are
$A_2^{(2)}$)\\
 & kernel doubling at $N=-2$ & \emph{explained}: the $X$--$Y$
couplings vanish, the densities decouple, the commutant factorizes\\
R7 & errata to \cite{Vitchev} & \emph{re-derived from scratch}, with
one stated exception: the cylindrical $\I_5$ eqs.~(54), (56), (58)
were not encoded, so the fifth erratum is \emph{presumed} there\\
R8 & corank $=$ graded character of the joint screening kernel &
\emph{conjecture}\\
\hline\hline
\end{tabular}
\caption{Claim status.  The results below mix all-spin theorems,
identities in the family parameter, interpolation proofs against a
degree bound, statements exact only at sampled parameter values,
finite-field certificates, and one conjecture.  This table records
which is which; where it is terse, the body is the authority.}
\label{tab:claims}
\end{table}

All computations are exact (rational, or rational-function)
arithmetic; engines were validated against one another, against the
Virasoro algebra, against known families, and with negative controls.
Code and machine-readable results accompany this paper.\footnote{%
Repository:
\texttt{https://github.com/evitchev/integrable-boundary-certificates},
tagged release \texttt{v1.1.5}.}

\section{Conventions}\label{sec:setup}

We follow \cite{Vitchev}. Chiral free bosons $X^j(z)$ carry the Wick
pairing $\langle\partial X^i(z)\,\partial X^j(w)\rangle
=\delta^{ij}(z-w)^{-2}$, hence
$\langle\partial^mX^i(z)\,\partial^nX^j(w)\rangle
=\delta^{ij}(-1)^{m-1}(m+n-1)!\,(z-w)^{-(m+n)}$. Densities are
normal-ordered polynomials in $\partial^mX^j$; a charge
$\I=\oint P\,dz$ vanishes iff $P$ is a total derivative, and
\[
[\textstyle\oint P,\oint Q]=\oint \operatorname*{Res}_{z\to w}
P(z)Q(w)\,dw ,
\]
so commutativity is the statement that the OPE residue lies in
$\im\partial$. Charges of spin $\sigma$ (density spin $\sigma+1$) form
graded spaces $\Q^\sigma$; the bracket adds charge spins. For a fixed
symmetry sector, expanding two candidate charges in bases of
$\Q^{\sigma_1}$, $\Q^{\sigma_2}$ turns $[\I_{\sigma_1},\I_{\sigma_2}]=0$
into a bilinear system for the coefficient pair --- the systems of
Section 2 of \cite{Vitchev}, which look overdetermined yet are
consistently solvable.

Three exact engines are used. (a)~A component engine over
$\mathbb{Q}$ (or $\mathbb{Q}(n)$) for explicit small $N$. (b)~An
$O(N)$-invariant engine: monomials are multisets of pairs
$(ab)=\partial^aX\!\cdot\!\partial^bX$; Wick contractions glue pairs
into index chains and loops, so residues stay in the invariant algebra
with coefficients in $\mathbb{Q}[N]$ --- $N$ is a \emph{formal}
parameter, and specializations are faithful whenever $N$ exceeds the
number of letters involved (first/second fundamental theorems of
$O(N)$ invariant theory). (c)~A mixed engine for
$\mathbb{Z}_2\times O(N)$: one distinguished boson $X$ plus $N$ bosons
entering through pairs. Engines (b), (c) were validated by exact
agreement with (a) at $N=1,2,3$ on random monomial pairs, and (a)
against the Virasoro OPE, translation invariance, antisymmetry mod
$\partial$, and a deliberately perturbed negative control.

\section{The bootstrap lemma and spectral rigidity}\label{sec:lemma}

\begin{lemma}\label{lem:boot}
Fix a distinguished odd-spin charge $D$ in a symmetry sector ---
$D=\I_3$ for the KdV, paperclip and cylindrical families, $D=\I_5$
for $A_2^{(2)}$, whose spins are $1,6m\pm1$ --- and suppose
\emph{(even-spin vacancy)} $\ker\ad_{D}\cap\,\Q^{\sigma}=0$ for all
even $\sigma$ in a range $E$. Choose any
$\I_\sigma\in\ker\ad_{D}\cap\,\Q^\sigma$ for odd $\sigma$. Then
$[\I_{\sigma_1},\I_{\sigma_2}]=0$ for all odd $\sigma_1,\sigma_2$ with
$\sigma_1+\sigma_2\in E$.
\end{lemma}

\begin{proof}
$J:=[\I_{\sigma_1},\I_{\sigma_2}]\in\Q^{\sigma_1+\sigma_2}$ has even
spin, and $[D,J]=[[D,\I_{\sigma_1}],\I_{\sigma_2}]
+[\I_{\sigma_1},[D,\I_{\sigma_2}]]=0$, so $J$ lies in an even-spin
kernel, which vanishes by hypothesis.
\end{proof}

Both forms of the lemma (the core step above and the odd-spin range
statement) are additionally certified by a kernel-checked Lean~4
proof in an arbitrary Lie algebra graded by submodules with
bracket-additive grading (\texttt{lean/Bootstrap.lean}, checked
against Mathlib with a fail-closed negative control; the
formalization also records that the distinguished element's own
grade never enters the hypotheses).  The same file kernel-checks the
degree-bound interpolation principle used in
Section~\ref{sec:paperclip}, and a companion
(\texttt{lean/Specialization.lean}) certifies the inference under
every $\mathbb{Q}(t)$ commutativity proof below: if the entrywise
evaluation of a polynomial matrix at a single parameter point has
trivial kernel, then so does the matrix itself over the polynomial
ring --- hence over $\mathbb{Q}(t)$ after the denominator clearing
our engines perform --- together with the bounded-degree
rational-section vanishing step.  (The formal statement is sharper than the prose route: the
one-point hypothesis yields the identical-vanishing conclusion
directly, with semicontinuity subsumed.)  The division of labor is
exact:
Lean certifies the abstract implications; the exact-arithmetic
engines certify the instance hypotheses (bracket-additivity of the
physical grading, the vacancy tables, the degree bounds) --- neither
replaces the other, and the free-boson realization itself is not
formalized.

The counting argument is classical in spirit
\cite{ZamCount}, and the odd/even Jacobi step itself is used inside
Feigin--Frenkel's proof of affine-Toda commutativity \cite{FF}; the
point here is to isolate it as a finite-range bootstrap criterion for
the bilinear systems of \cite{Vitchev} and
verify the hypotheses by machine. Consequences: the only nonlinear step in the
bilinear systems of \cite{Vitchev} is the choice of the distinguished
charge $D$ (the distinct families there are distinct $D$ branches);
each higher charge
solves the \emph{linear} problem $[D,\I_\sigma]=0$; and every
remaining bilinear equation is an identity within the verified
vacancy range. Geometrically, the
commuting variety cut out by the bidegree-$(1,1)$ equations fibers
over the determinantal locus of admissible $D$'s with linear
fibers (a description verified in our finite instances, not a global
scheme-theoretic theorem) --- the observed
excess intersection is graded Lie theory, not accident.

\begin{table}[htbp]\centering
\begin{tabular}{lccccccccc}
\hline\hline
charge spin $\sigma$ & 2 & 3 & 4 & 5 & 6 & 7 & 8 & 9 & 10\\
\hline
KdV ($N{=}1$): $\dim\Q^\sigma$ & 0&2&0&3&1&5&2&8&5\\
KdV: $\dim\ker\ad_{\I_3}$ & 0&1&0&1&\textbf{0}&1&\textbf{0}&1&\textbf{0}\\
paperclip: $\dim\Q^\sigma$ & 0&5&1&11&7&25&21&55&57\\
paperclip: $\dim\ker\ad_{\I_3}$ & 0&1&0&1&\textbf{0}&1&\textbf{0}&1&\textbf{0}\\
\hline\hline
\end{tabular}
\caption{Genuine (modulo total derivatives) kernels of $\ad_{\I_3}$.
Paperclip values at five generic rational $n$; vacancy at one point
implies generic vacancy. Note the nontrivial rigidity: at $\sigma=10$
a 57-dimensional space carries exactly zero kernel.}
\label{tab:spectral}
\end{table}

Table~\ref{tab:spectral} verifies both hypotheses --- even-spin
vacancy and odd-spin multiplicity one --- for the $c=1$ KdV branch and
the paperclip branch. Multiplicity one means the towers are
\emph{rigid}: given $\I_3$ there are no deformations of higher
charges. Combined with Lemma~\ref{lem:boot} and the symbolic
verifications below, $\{\I_1,\I_3,\I_5,\I_7\}$ of the paperclip
family pairwise commute for \emph{all} $n$ away from finitely many
poles.

\section{The paperclip family: $P_8$, exact commutativity,
screenings}\label{sec:paperclip}

\subsection{Corrected $\I_5$ and the new $\I_7$}

Solving $\ker\ad_{\I_3}$ at spin 5 from scratch at $n=2,3,7/3$
uniquely reproduces eq.~(51) of \cite{Vitchev} except for one
coefficient: the $P_6^{[10]}=(111111|)$ term must read
$-\tfrac{n(6+5n)(8+5n)}{120(2+n)}$ (the print omits the factor $n$;
the two agree only at $n=1$). With this correction,
$[\I_3,\I_5]=0$ holds identically in $n$ (symbolic check).

At spin 7 the kernel is one-dimensional at each of 39 sampled rational
$n$; rational-function reconstruction with held-out verification, and
a final symbolic check of $[\I_3,\I_7]=0$ identically in $n$, give the
closed form of $P_8$ listed in Appendix~\ref{app:P8}. Structural
features: common denominator $15\,n^3(7n{+}8)(7n{+}10)(7n{+}12)$;
covariance under $n\to-2-n$ (the $X\leftrightarrow Y$ duality of the
paperclip), which exchanges the factor sets $\{7n{+}2,7n{+}4,7n{+}6\}$
and $\{7n{+}12,7n{+}10,7n{+}8\}$; denominator factors $(7n{+}8)(7n{+}10)(7n{+}12)$ and numerator factors
$(7n{+}2)(7n{+}4)(7n{+}6)$ --- together the six resonance factors of
the quasiclassical Wronskian coefficient $S_7$ of the paperclip ODE
(eq.~(68) of \cite{Vitchev}) ---
continuing the pattern $(3n+2j)$ in $P_4/S_3$ and $(5n+2j)$ in
$P_6/S_5$: the analytic structure of the spin-$(2k{+}1)$ density
mirrors the spectral data of the conjectured oper at every computed
order. Finally $[\I_5,\I_7]=0$ was verified at 20 rational
$n$; after clearing denominators every entry of the reduced commutator
is a polynomial of degree $\le14$ in $n$, so the identity holds for
all $n$. A unique $\I_9$ exists at each sampled generic $n$.

\subsection{Screening vectors and the spin-7 test}

Demanding that a vertex charge $\oint e^{\alpha\cdot X}$ commute with
$\I_3$ and $\I_5$ (a twisted total-derivative condition in the
Feigin--Frenkel sense) returns the four vectors
\[
\alpha=\Bigl(\pm\sqrt{-n/2},\;\pm\sqrt{(n+2)/2}\Bigr),
\qquad \alpha_1^2+\alpha_2^2=1 ,
\]
the known screenings $\exp(\sqrt nX\pm i\sqrt{n+2}\,Y)$ of \cite{LVZ}
after the normalization map of \cite{Vitchev}.  (In this bulk
normalization $|\alpha|^2=1$ gives chiral weight $\tfrac12$,
i.e.\ boundary marginality $2h=1$: these are boundary perturbing
exponentials in the sense of \cite{LVZ}, and ``screenings'' is
used for them throughout in that boundary sense.) (Calibration: the same
fit on the $c=1$ KdV branch returns $\alpha=\pm\sqrt2$, the $SU(2)$
point.) The decisive test: all four screenings annihilate the
independently computed $\oint P_8$ --- a genuine prediction, as $P_8$
entered nowhere in the fit --- verified in exact arithmetic over
$\mathbb{Q}(i,\sqrt2)$-type fields at $n=2,3$ (at $n=2$ the fit also
returns two extra pure-$Y$ vectors; see R4). At the sampled $n$, this verifies the \emph{annihilation} half of
the four-screening characterization of the paperclip tower
(eq.~(84) of \cite{LVZ}) beyond the low spins checked in 2004; the
\emph{uniqueness} half --- that the joint kernel of the four
conditions on the $\mathbb{Z}_2\times\mathbb{Z}_2$-invariant sector
is one-dimensional per spin --- is certified separately at
the same samples (spins 7 and 9, $n=2,3$).  Uniqueness of the joint
kernel for the four \emph{known} vectors is distinct from
completeness of the screening-vector \emph{locus}, which remains
scoped as in R4.

\section{Symbolic $N$: the $A_2^{(2)}$ family and the branch-point
mechanism}\label{sec:a2}

\subsection{All-spin identification}

That the $\Phi_{1,2}$ and $\Phi_{2,1}$ perturbations of Virasoro
carry the $A_2^{(2)}$ (Zhiber--Mikhailov--Shabat
\cite{ZhiberShabat,MikhailovRed}) integrable
structure, with conserved spins governed by the affine exponents,
is classical \cite{ZamCount,Smirnov}; the conformal-perturbation
route from CFT to the integrable hierarchies goes back to the same
work and to \cite{EguchiYang,SasakiYamanaka}; what follows is not that
identification but an exact, certified dictionary attaching the
specific two-sheeted family of \cite{Vitchev} to it, in the
normalization of that paper.  With the dictionary in place, the
$A_2^{(2)}$ label is promoted from a spin-content diagnosis to
an all-spin statement.  Introduce a Coulomb-gas parameter $a$ and
background charge \cite{DF84,DF85}
\[
 Q=\frac{a}{2}-\frac{1}{a},\qquad
 T=\frac{1}{2}:(\partial X)^2:+Q\,\partial^2X .
\]
In our OPE convention this stress tensor has
\[
 c=1-12Q^2=13-3a^2-\frac{12}{a^2},\qquad
 s=3a^2-\frac{12}{a^2},\qquad
 s^2=(c-25)(c-1).
\]
(Sign convention: at $a^2=\tfrac12$, the $c_{1,4}=-\tfrac{25}{2}$
point, choosing $a=+\tfrac{1}{\sqrt2}$ gives $Q=-\tfrac{3}{2\sqrt2}$, while
Section~\ref{sec:c1} works with the opposite sign,
$Q=+\tfrac{3}{2\sqrt2}$, related by $X\to-X$, under which the
screening momenta map accordingly.)
An exact point-split OPE calculation, including the twisted
total-derivative witness, gives on the generic stratum
\[
\begin{split}
 \left[\oint e^{X/a},\oint\bigl(:T^3:+A:(\partial T)^2:\bigr)\right]=0
 &\quad\Longleftrightarrow\quad
 A=\frac{12a^4-5a^2+3}{8a^2},\\
 \left[\oint e^{-aX/2},\oint\bigl(:T^3:+A:(\partial T)^2:\bigr)\right]=0
 &\quad\Longleftrightarrow\quad
 A=\frac{3a^4-20a^2+192}{32a^2}.
\end{split}
\]
On the other hand, eq.~(36) of \cite{Vitchev}, normalized to unit
coefficient of $:T^3:$, has
\[
 A_\pm=\frac{181-17c\mp15s}{64}.
\]
Substitution of the preceding parametrization identifies $A_-$ with
the first screened expression and $A_+$ with the second.  Thus the
sheets of the density of \cite{Vitchev} are exactly the dual $\Phi_{1,2}$ and
$\Phi_{2,1}$ Coulomb-gas
realizations.  The symbolic certificate is
\texttt{code/a2\_infinitude\_certificate.py}.

One further lemma connects these single-screening statements to the
affine-Toda formalism.  The two ordinary Virasoro screenings
$\oint e^{aX}$ and $\oint e^{-2X/a}$ \cite{FeiginFuchs} impose
\emph{no} condition on the pencil.  Indeed both exponentials have
$T_Q$-weight one at the background charge fixed by $a$, so each is a
weight-one Virasoro primary; the contour integral of a weight-one
primary commutes with every Virasoro mode, hence with $T$ itself,
hence with every normally ordered differential polynomial in $T$ and,
after integration, with every $T$-composite charge.  Every member of
the pencil $:T^3:+A:(\partial T)^2:$ is such a charge, so both
ordinary screenings annihilate the whole pencil identically in $A$ and
select nothing.  The surviving condition is therefore the perturbing
one, and each displayed equation places the corresponding sheet in the
full twisted affine-Toda joint kernel, $\{e^{aX},e^{-aX/2}\}$ for one
sheet and the dual pair $\{e^{-2X/a},e^{X/a}\}$ for the other (root
ratio $-\tfrac12$ in both; sheet exchange is the coupling duality
$a\mapsto-2/a$).  On the generic stratum the perturbing condition cuts
the two-dimensional weight-six pencil down to the single line
displayed above, so the joint kernel there is one-dimensional and the
identification of $\I_5$ with the Feigin--Frenkel generator is
forced rather than merely consistent.  The certificate asserts the
same statement computationally.

Feigin and Frenkel's Theorem~4.7.3 \cite{FF} --- proved there for
affine algebras whose exponents are odd and whose Coxeter number is
even, a case that includes $A_2^{(2)}$ (exponents $6n\pm1$, $h=6$),
with the general case deferred --- proves that, for generic coupling,
the classical affine-Toda integrals deform to mutually commuting
quantum local integrals in the affine-exponent degrees.  Their
$A_2^{(2)}$ discussion identifies the resulting Virasoro conservation
laws with the $\Phi_{1,2}$ perturbation; the degrees are precisely the
positive integers not divisible by $2$ or $3$.  Combining that theorem
with the exact dictionary above proves that, over the generic point of
$s^2=(N-25)(N-1)$, the family containing the $I_5$ of \cite{Vitchev} is infinite,
with a commuting generator at every charge spin $6m\pm1$.  This is a
generic theorem: at resonant central charges the cohomology and the
irreducible quotient can jump, so special fibers must still be treated
separately.

\subsection{The Ising specialization: a certified degeneration}

The genericity caveat above is not hypothetical, and its first
nontrivial instance is now exhibited exactly.  At $t=\pm7$ the
parameter curve meets the Ising central charge:
$N=c=\tfrac12$, $s=\mp\tfrac72$.  The Ising vacuum module has its
first nontrivial singular vector at level $(3-1)(4-1)=6$ ---
precisely the weight of the spin-5 density --- and the certificate
\texttt{ff\_ising\_quotient.py} derives that null field in-engine
as the unique primary weight-6 $T$-composite (present at both dual
Coulomb-gas realizations $a^2=\tfrac32,\tfrac83$; absent at a
control point whose central charge admits no level-6 vacuum null).
Writing the two sheets of the spin-5 density as
$8\,{:}T^3{:}+K_s\,{:}(\partial T)^2{:}$, the certified verdict is
one-sided: the $s=-\tfrac72$ sheet ($t=+7$, $K=\tfrac{225}{8}$)
\emph{equals the null-field class modulo total derivatives} and
therefore vanishes in the irreducible quotient, while the
deck-conjugate $s=+\tfrac72$ sheet ($t=-7$, $K=15$) survives as a
nonzero class --- verified with exact rank certificates at both
realizations.  (Labeling note: $K=\tfrac{225}{8}$ at $s=-\tfrac72$ is
$8A_+$ in the pencil normalization --- equivalently $8A_-$ at
$s=+\tfrac72$; Section~\ref{sec:a2} attaches $A_-$ to the first
screened realization, so tracking sheets by the sign of $s$ alone
inverts the naming.)  In Ising terms the two sheets are the two relevant
directions: the vanishing $K=\tfrac{225}{8}$ sheet is the
$\Phi_{1,2}$ (magnetic, perturbing weight $h=\tfrac1{16}$, $E_8$)
realization, the surviving $K=15$ sheet the $\Phi_{2,1}$
(energy/free-fermion) one --- the certified counterpart of the
classical absence of a spin-5 charge in Ising in a magnetic field
\cite{ZamIsing}
(the sheet$\leftrightarrow$operator naming is interpretation from
the perturbing-weight arithmetic, on top of the certified
$K$-values and ranks).
The degree-5 dropout itself is prior art
(\cite{FF}, \S4.8.4, citing its discovery in earlier work
\cite{ZamCount}; singular vectors as probes of integrable
perturbations are developed systematically in \cite{MathieuWatts});
the
explicit density/null-class mechanism in the present normalization
and the one-sheet resolution across the deck transformation
$s\mapsto-s$ are, to our knowledge, contributions of this
certification.  The
degeneration is exactly the specialization behavior the generic
theorem above permits, exhibited at its first rational instance.

\subsection{Certification of the $\I_7$ formulas}

Solving $\ker\ad$ at spin 7 from scratch in the invariant algebra and
reconstructing in $N$ (and in $s=\sq$, split over the two sheets)
certifies: eq.~(33) of \cite{Vitchev} exactly, including its $N^3$
term; eq.~(30) up to the misprint $\tfrac2{657}\to\tfrac2{675}$ ---
forced independently by the coincidence of the KdV and $O(N)$ families
at $N=1$; and eq.~(35) up to two dropped minus signs
(Appendix~\ref{app:errata}). Eqs.~(36)--(37) with the composite
expansions (43)--(45) are correct as printed.

\subsection{Generic-$N$ spin content and the $c=1$ enhancement}

On the rational parametrization $N=(t^2-25)/(t^2-1)$,
$s=-24t/(t^2-1)$ of the double cover (branch points $t=0\mapsto N=25$
and $t=\infty\mapsto N=1$; deck transformation $t\to-t$), the genuine
kernel of $\ad_{\I_5^{A_2}}$ over charge spins $\sigma=2,\dots,13$
equals
\[
0\;0\;0\;1\;0\;1\;0\;0\;0\;1\;0\;1
\]
at five generic points (both sheets), \emph{and also} at the branch
point $N=25$ --- where the formal invariant algebra is faithful for
these letter counts --- \emph{and} for the branch coefficients
$(N,s)=(1,0)$ evaluated in the formal algebra. Thus: the spin content
is pure $6n\pm1$ (the exponents of $A_2^{(2)}$), with the members at
spins 11 and 13 subsequently constructed over $\mathbb{Q}(t)$
(generic existence proven); vacancy at spins 3 and 9 is generic by rank
semicontinuity on the irreducible parameter curve; and the known
enhancement at physical $N=1$ --- where an all-odd pattern appears
through the verified range,
containing $\I_3'=\oint[(\partial X)^4-\tfrac52\partial X\partial^3X]$
with $[\I_3',\I_7]=[\I_3',\I_9]=[\I_7,\I_9]=0$ and
$[\I_3',\I_3^{\rm KdV}]\neq0$ --- is, through the verified range, driven at the
invariant-algebra level by the finite-$N$ relations (e.g.
$(22)(11)=(12)(12)$ at $N=1$).  Section~\ref{sec:c1} resolves the
structural question: the tower is the quantum KdV hierarchy at
$c_{1,4}=-\tfrac{25}{2}$ in Feigin--Fuchs realization (the all-odd
quantum-KdV sector of the $S_4$ family of \cite{FHW}, whose first
odd-parity $\Delta$-classes are realized in the computed
odd sector of Section~\ref{sec:mechanism}, with all-spin exhaustiveness open; the embedding is what
is new), the
mechanism a $\mathbb{Z}_2$ symmetrization of the twisted screening
set, available exactly where the background charge vanishes ---
pinned, after all, to the self-dual branch point, with the
finite-$N$ relations its invariant-algebra face; the same mechanism
explains why $N=25$ shows no enhancement ($Q^2=-2$ there).  Fibers
of the commuting-tower family jump at degeneration loci of the
invariant algebra; at $N=1$ the locus and the branch point coincide
because both express the vanishing of the background charge.
Within the maximal-commutative-subalgebra menagerie of the
Bakas--Kiritsis-type $W_\infty$ \cite{BK,KL}, the answer is thus
KdV itself at shifted coupling, not a new member. As a by-product the
graded dimensions of the $O(N)$-invariant charge spaces reproduce
Table~1 of \cite{Vitchev} and extend it: $\dim\Q^{13}=59$.

\section{The $c=1$ fiber: an exact identification}\label{sec:c1}

Section~\ref{sec:a2} left the $c=1$ enhancement diagnosed but not
explained: at the branch point the centralizer acquires a charge at
every odd spin, and the finite-$N$ relations of the invariant algebra
account for why at the level of that algebra, but say nothing about
what the enhanced tower \emph{is}.  This section identifies it, in the
one-boson realization where the question is sharp.

\subsection{The enhanced tower}\label{sec:object}

In the conventions of Section~\ref{sec:setup} specialized to a
single chiral boson $\varphi$, the $A_2^{(2)}$ family reaches at its
branch point $c=1$, where its spin-5 charge has density
\begin{equation}\label{eq:I5}
P_6=\tfrac{31}{24}\,\partial^3\varphi\,\partial^3\varphi
+\tfrac52\,(\partial\varphi)^2(\partial^2\varphi)^2
+(\partial\varphi)^6
\pmod{\operatorname{Im}\partial}.
\end{equation}
Charges split under the parity $\varphi\mapsto-\varphi$ into
$\mathbb{Z}_2$-even and $\mathbb{Z}_2$-odd sectors; throughout,
\emph{the tower} means the $\mathbb{Z}_2$-\emph{even} part of the
centralizer $\ker\ad_{\I_5}$ (the odd part is nonempty and is
described in Section~\ref{sec:mechanism}).  The tower at this
point is \emph{enhanced}:
through the scanned range $\sigma\le10$ (i.e.\ odd spins
through $9$) it contains a charge at
every odd spin, each with multiplicity one.  Its spin-3 member is
the ray
\begin{equation}\label{eq:I3p}
\I_3'=\oint\Big[(\partial\varphi)^4
-\tfrac52\,\partial\varphi\,\partial^3\varphi\Big]
\;\equiv\;\oint\Big[(\partial\varphi)^4
+\tfrac52\,(\partial^2\varphi)^2\Big],
\end{equation}
which is \emph{not} the $c=1$ quantum KdV ray
$(\partial\varphi)^4-2(\partial^2\varphi)^2$: the rays differ
modulo $\partial$, and $[\I_3^{\rm KdV},\I_5]$ is certified not to
be a total derivative.  It is, however --- and this is the central
point of this section --- exactly the KdV ray at a \emph{shifted
background charge} on the same boson
(Section~\ref{sec:ident}).  The charges $\I_3',\I_5,\I_7',\I_9'$
pairwise commute in the verified range.

Two disclaimers govern everything below.  First, all-spin
\emph{existence} of the tower is proven in
Section~\ref{sec:mechanism}; all-spin \emph{multiplicity one} is
not --- per-spin uniqueness is certified through $\sigma\le9$ and
independently verified through $\sigma\le15$, and remains
finite-range beyond.  Second, ``certified'' has throughout this paper the
meaning fixed in Section~\ref{sec:corank} under Methods.

\subsection{The identification}\label{sec:ident}

Introduce on the \emph{same} boson the Feigin--Fuchs stress tensor
\[
T_Q=\tfrac12\,(\partial\varphi)^2+Q\,\partial^2\varphi,
\qquad c(Q)=1-12Q^2 ,
\]
and recall that for each $Q$ the quantum KdV hierarchy
\cite{BLZ1,BLZ2,BLZ3,FF} is the commuting family of local charges built from
$T_Q$; only $Q$ selects which densities are distinguished --- the
underlying OPE bracket is that of the single boson throughout.
Throughout, $:AB:$ denotes the point-split (Wick) composite --- the
$O(1)$ term of the $A(z)B(w)$ OPE; its double-contraction terms are
what shift the naive coefficient $Q^2$ in identity~\emph{(i)} below
to $Q^2-\tfrac12$ (in particular, at $Q=0$ the composite is
$:T^2:\;\equiv\;\tfrac14(\partial\varphi)^4
-\tfrac12(\partial^2\varphi)^2$, not the naive square).

\begin{theorem}[certified operator identities]\label{thm:ident}
\emph{(i)} Symbolically in $Q$,
\[
:T_Q\,T_Q:\;\equiv\;
\tfrac14\Big[(\partial\varphi)^4+(4Q^2-2)(\partial^2\varphi)^2\Big]
\pmod{\operatorname{Im}\partial}.
\]
Hence the enhanced ray \eqref{eq:I3p} equals $4\oint:T_Q^2:$
exactly when $4Q^2-2=\tfrac52$, i.e.
\[
Q^2=\tfrac98,\qquad
c=1-12Q^2=-\tfrac{25}{2}=c_{1,4}.
\]
\emph{(ii)} At $Q^2=\tfrac98$,
\[
:T_Q\!:\!T_QT_Q\!:\!:\;+\;\tfrac78\,:\partial T_Q\,\partial T_Q:
\;\equiv\;\tfrac18\,P_6
\pmod{\operatorname{Im}\partial},
\]
with the coefficient $\tfrac78=(4Q^2-1)/4$ the standard KdV value
fixed by $[\,\oint:T_Q^2:,\,\cdot\,]=0$.  Hence the tower's
defining charge $\oint P_6$ is the quantum KdV $\I_5$ at
$c=-\tfrac{25}{2}$.
\end{theorem}

Both identities are certified symbolically (exact arithmetic; the
first identically in $Q$, with negative controls rejecting the
wrong quantum shift and the wrong specialization point).
Identity \emph{(i)} is not new: composite quantum-KdV identities
of this kind go back to \cite{KM}, and with the dictionary
$g=-\tfrac13(c+5)$ it appears as eqs.~(2.14)--(2.15) of
Freeman--Hornfeck--West \cite{FHW}, whose one-boson commuting
series $S_4$ (lowest odd current of weight $h=7$, coupling
$g=\tfrac52$, $c_4=13-6\cdot4-\tfrac64=-\tfrac{25}{2}$) is
precisely the $c_{1,4}$ structure at issue.  What is new here is
the \emph{embedding}: the $c=1$ branch point of the $A_2^{(2)}$
family lands on that known structure.  Freeman--Hornfeck--West
remark that the full spin content of their series matches the
exponents of no Lie algebra.  The embedding does not alter that
observation; rather, it identifies the $6n\pm1$ $A_2^{(2)}$
subhierarchy inside the all-odd KdV sector of $S_4$ and explains
why the centralizer enlarges at the branch point.  Since the
KdV charges at fixed $Q$ mutually commute \cite{BLZ1,FF}, the
entire hierarchy at $Q^2=\tfrac98$ lies in $\ker\ad_{\I_5}$; the
certified multiplicity one then forces class-by-class equality
wherever both are nonzero.  Concretely, the even-sector $\ker\ad_{\I_3'}$ is
certified to have multiplicity exactly one at $\sigma=5,7,9$ and
zero at even spins, with its class \emph{equal} to the tower class
at each odd spin; an independent engine (see Acknowledgments)
extends the even-sector coincidence
$\ker\ad_{\I_5}=\ker\ad_{\I_3'}$
class-by-class through $\sigma=15$, with spin 14 vacant (the
full-space centralizer carries, in addition, the odd-parity
classes of Section~\ref{sec:mechanism}).  Within the certified
range the tower \emph{is} the quantum KdV hierarchy at $c_{1,4}$
in Feigin--Fuchs realization on the $c=1$ boson; at every spin it
contains it (Theorem~\ref{thm:inf}).

\subsection{Mechanism and infinitude}\label{sec:mechanism}

Why does the twisted $A_2^{(2)}$ structure, whose exponents are
$6n\pm1$, sit inside an all-odd-spin KdV tower precisely at the
branch point?  The answer is a two-line parity argument, certified
at the level of kernels.

\begin{lemma}[$\mathbb{Z}_2$ symmetrization]\label{lem:z2}
Let $P$ be even under $\varphi\mapsto-\varphi$.  Then the
first-order conservation condition of $\oint P$ under
$e^{-\alpha\varphi}$ is the parity image of the condition under
$e^{+\alpha\varphi}$; on the even sector the two conditions have
identical kernels.
\end{lemma}

\begin{proof}
Parity $\varphi\mapsto-\varphi$ is an automorphism of the OPE
sending $V_\alpha\mapsto V_{-\alpha}$ and the twisted derivative
$\partial+\alpha\,\partial\varphi$ to
$\partial-\alpha\,\partial\varphi$; hence the $-\alpha$
condition on $P$ is the $+\alpha$ condition on the parity image of
$P$, and the kernels coincide sector by sector.
\end{proof}

The lemma is elementary, and the parity argument is classical ---
it appears in \cite{FHW}, there credited to Sasaki--Yamanaka, in
the $S$-series context; its content is nonetheless certified as
kernel coincidence (three momenta, density weights $2$--$9$), with
a control confirming that off the even sector the $\pm\alpha$
kernels genuinely differ.  At the branch point the background
charge vanishes, so every charge of the family --- and every
$T$-composite --- is $\mathbb{Z}_2$-even.  The $c=1$ specialization
of the $A_2^{(2)}$ screening set is the pair
$\{e^{\sqrt2\varphi},\,e^{-\varphi/\sqrt2}\}$; by the lemma its
even-sector joint kernel equals that of the symmetrized quartet
$\{e^{\pm\sqrt2\varphi},\,e^{\pm\varphi/\sqrt2}\}$ (certified at
every weight $2$--$9$), and in particular is contained in the joint
kernel of the sine-Gordon pair $\{e^{\pm\varphi/\sqrt2}\}$.  That
pair is the $A_1^{(1)}$ affine Toda screening set at coupling
$a=\tfrac1{\sqrt2}$ (``screening'' in the affine-Toda sense ---
the simple-root exponentials of $A_1^{(1)}$; under the $c_{1,4}$
Virasoro grading of Section~\ref{sec:fingerprint} one member of
the pair is a weight-one screening and its mirror a
$\Phi_{1,3}$-type perturbing operator), whose joint kernel
\emph{contains} the quantum KdV/sine-Gordon hierarchy at that
coupling and equals it for generic coupling \cite{FF,BLZ1}.  Thus
$A_2^{(2)}|_{c=1}$ lands in the sine-Gordon pair kernel at
$a=\tfrac1{\sqrt2}$; through the certified range that kernel is
exactly the hierarchy of Section~\ref{sec:ident}, so
\[
A_2^{(2)}\big|_{c=1}\;\subset\;
\mathrm{KdV}_{c=-25/2}
\quad\text{(within the certified range),}
\]
and since the spin-5 charges coincide, $\ker\ad_{\I_5}$ swallows
the whole all-odd tower: the $6n\pm1$ exponents of the twisted
family sit inside the odd integers of $A_1^{(1)}$, and the
``enhancement'' is the difference between the two exponent sets.
Away from $c=1$ the $T_Q$-composites are not $\mathbb{Z}_2$-even
and the symmetrization fails --- which is why the generic
$A_2^{(2)}$ content is only $6n\pm1$, and why the enhancement is
pinned to the branch point.

\begin{lemma}[primitive specialization]\label{lem:prim}
Let $R=\Q[a,a^{-1}]$, let $M$ be a matrix over $R$, and let
$v\in R^{\,n}$ be a nonzero vector with $Mv=0$.  Then there is a
$w\in R^{\,n}$ with $Mw=0$ such that the specialization $w(a_0)$ is a
\emph{nonzero} kernel vector of $M(a_0)$ for \emph{every}
$a_0\in\mathbb{C}^{\times}$ (evaluation is a ring homomorphism
$R\to\mathbb{C}$; only $a_0=0$ is excluded) --- in particular at the
irrational point $a_0=\tfrac1{\sqrt2}$ used below.
\end{lemma}

\begin{proof}
$R$ is the localization of the principal ideal domain $\Q[a]$ at
$\{a^{k}\}$, hence is itself a principal ideal domain, with units the
monomials $\lambda a^{k}$, $\lambda\in\Q^{\times}$.  Put
$d=\gcd(v_1,\dots,v_n)\in R\setminus\{0\}$ and $w=v/d\in R^{\,n}$, so
that $w$ is \emph{primitive}: $(w_1,\dots,w_n)=R$.  Since $R$ is a
domain and $d\neq0$, $Mv=d\,Mw=0$ gives $Mw=0$.  Primitivity supplies
$u_i\in R$ with $\sum_i u_iw_i=1$ (B\'ezout).  Every element of $R$ is
regular at each $a_0\neq0$, so evaluation there is a ring homomorphism
$R\to\mathbb{C}$; applying it to that identity gives
$\sum_i u_i(a_0)w_i(a_0)=1$, whence $w_i(a_0)\neq0$ for some $i$.
Specializing $Mw=0$ at $a_0$ gives $M(a_0)w(a_0)=0$.
\end{proof}

The Laurent ring is what makes the conclusion uniform in $a_0\neq0$:
$a$ is a unit in $R$, so the normalization $w=v/\gcd$ absorbs any
power of $a$ that the coordinates share, and no further point of the
line is lost.  The single excluded point $a_0=0$ is excluded already
in the theorem below, and is not an artifact of the normalization:
evaluation at $0$ is not defined on $R$.  (If the coordinates happen
to lie in $\Q[a]$, the same B\'ezout argument runs over $\Q[a]$ and
excludes nothing; only the weaker statement is needed.)  Note the
direction: Lemma~\ref{lem:prim} propagates a \emph{nonzero} class
\emph{down} to a special point, whereas the specialization lemma
kernel-checked in \texttt{lean/Specialization.lean} propagates
\emph{vacancy} \emph{up} from one point to the whole line.  The two
are distinct statements and neither implies the other; only the
latter is formalized.

\begin{theorem}[all-spin existence]\label{thm:inf}
For every $a\neq0$ the even-sector joint kernel of
$\{e^{\pm a\varphi}\}$ contains at least one charge at every odd
spin.  In particular the enhanced tower contains a charge at every
odd spin.
\end{theorem}

\begin{proof}
The exponents of $A_1^{(1)}$ are the odd integers, so by
Feigin--Frenkel \cite{FF} the joint kernel has exactly one charge
per odd spin at generic $a$.  The generic kernel line is moreover
$\mathbb{Z}_2$-\emph{even}: the pair $\{\pm a\}$ of conditions is
parity-stable, so the one-dimensional kernel is spanned by a
parity eigenvector, and the eigenvalue is even because the
charge's classical limit is the classical Toda density with
leading term $(\partial\varphi)^{\sigma+1}$ --- an even number of
factors at odd $\sigma$; evenness persists identically in $a$
(the odd-monomial coordinates of the $\mathbb{Q}[a]$ class vector
vanish at generic $a$, hence identically), so the specialized
charges below land in the even sector, i.e.\ the tower.
At fixed spin the conditions form a finite linear system whose
matrices depend polynomially on $a$ (the twisted image varies with
$a$; for $a\neq0$ the source and quotient dimensions are
constant: the source and the ordinary total-derivative subspace are
$a$-independent, and the twisted derivative at momentum $\pm a\neq0$
is injective on each twisted weight space --- a nonzero momentum
admits no closed twisted densities: in the letter-count filtration the
top component of $(\partial+a\alpha\,\partial\varphi)S=0$ reads
$\partial\varphi\cdot S_{\mathrm{top}}=0$, forcing $S=0$ --- so the
twisted-image dimension
is the same for every $a\neq0$, jumping only at the excluded
$a=0$); the generic rank is maximal on the parameter line, so
the kernel dimension can only grow under specialization: at least
one at every $a\neq0$, in particular at $a=\tfrac1{\sqrt2}$.  To
place charges in the tower, specialize the generic family itself:
the $A_1^{(1)}$ Toda integrals admit $\mathbb{Q}[a]$-primitive
\emph{class vectors} --- coordinates in a fixed, $a$-independent
basis of $\Q^\sigma$, so that primitivity is meant modulo total
derivatives (primitivity of a density representative would not
suffice: a density nonzero at the special point can still be a
total derivative there; existence is elementary: at each odd spin
the kernel over $\mathbb{Q}(a)$ is one-dimensional, so any nonzero
solution, with denominators cleared and the gcd of its coordinates
divided out, is a $\mathbb{Q}[a]$-primitive class vector, unique up
to a rational unit; that the screening condition is represented by a
matrix over $R$ on the fixed class coordinates follows from the
injectivity above by Smith normal form: over the PID $R$ the twisted
derivative's invariant factors are units --- it is injective at every
$a\neq0$ --- so its cokernel is a free $R$-module and ``residue lies
in the twisted image'' is a genuine $R$-matrix condition) --- whose
pairwise commutators are total
derivatives \emph{identically in} $a$ (they vanish at generic $a$
\cite{FF,BLZ1}, and a polynomial identity holding generically
holds identically); specializing at $a=\tfrac1{\sqrt2}$ --- by
Lemma~\ref{lem:prim} the specialized class is nonzero --- yields
charges commuting with the specialized $\I_5$; since the even-sector
kernel at spin 5 is one-dimensional at $a=\tfrac1{\sqrt2}$ (certified
below) and Theorem~\ref{thm:ident}\emph{(ii)} places $\oint P_6$ on
that line, the specialized $\I_5$ spans the same ray as $\oint P_6$.
Hence the tower contains a charge at every odd spin.
\end{proof}

In the certified range the even-sector dimension is exactly one at
every odd spin $1$--$9$ (with a genericity control at $a=2$), so
nothing degenerates at the special coupling through that range.
That such degenerations are \emph{possible} at $c_{1,p}$ couplings
is a certified fact: at $a=1$ (the $c_{1,2}=-2$ point) the
even-sector pair kernel has dimension \emph{two} at spin 7.
All-spin uniqueness statements are therefore deliberately left
finite-range.

The first structure of the centralizer's $\mathbb{Z}_2$-\emph{odd}
sector is now also identified (through the verified range; a complete
description is open problem~(ii) below).  The full-space $\ker\ad_{\I_5}$ contains the
odd-parity spin-6 charge
\[
J=\oint\Big[\tfrac{64}{7}(\partial\varphi)^7
-40\,(\partial\varphi)^3(\partial^2\varphi)^2
+\partial\varphi\,(\partial^3\varphi)^2\Big],
\]
certified to be a nonzero class commuting with
$\I_3',\I_5,\I_7',\I_9'$, conserved under $e^{\pm\varphi/\sqrt2}$
and not under $e^{\pm2\sqrt2\varphi}$.  Behind this pattern is an
operator identity: with
$Q_{\pm2\sqrt2}=\oint e^{\pm2\sqrt2\varphi}$,
\[
[Q_{2\sqrt2},\,Q_{-2\sqrt2}]\;\equiv\;\tfrac{\sqrt2}{45}\,J
\pmod{\operatorname{Im}\partial},
\]
with the parity-even part of the commutator residue a total
derivative, as parity requires.  The identity is classical: the
commutator of the screening charge with the opposite exponential
is precisely how Freeman--Hornfeck--West construct the first
odd-spin current of their $S_n$ series (\cite{FHW},
eqs.~(4.5)--(4.6)); at $n=4$ its charge is the identity above,
and the current is the $h_{3,1}=7$ generator of the $c_{1,4}$
singlet algebra --- their $\mathsf{W}(7)$, which is the $p=4$ member
$\mathsf{W}(2,2p-1)$ of the free-boson subalgebras introduced by
Kausch \cite{Kausch91}, whose irreducible
modules were classified by Adamovi\'c \cite{Adamovic} (logarithmic
intertwining structure in \cite{AdamovicMilas}), at $c=1-6(p-1)^2/p=-\tfrac{25}{2}$; the
algebra and its generator are prior art, and what is new here is the
exact normalization and the placement of the generator and its
commutator descendants inside this $c=1$ centralizer.  It was independently
rediscovered by the second-round review (see Acknowledgments) and
is certified, with the exact constant, in this project's engine.
(Grading dictionary, to prevent a translation trap: current weight
$=$ charge spin $+1$ throughout.  FHW's \emph{even}-weight currents
--- weights $4,6,8,\dots$ --- integrate to this paper's \emph{odd}
charge spins $3,5,7,\dots$, the $\mathbb{Z}_2$-\emph{even} tower;
FHW's \emph{odd}-weight currents --- weights $7,13,\dots$ ---
integrate to \emph{even} charge spins $6,12,\dots$, the
$\mathbb{Z}_2$-\emph{odd} sector discussed here.)
By the Jacobi identity, $J$ therefore commutes with \emph{every}
charge annihilated by both $\pm2\sqrt2$ conditions: the entire
quartet kernel, at all spins.  The same review finds one further
odd class at spin 12 --- the nested commutator
$[[Q_{2\sqrt2},J],Q_{-2\sqrt2}]$, an instance of the
$\Delta$-operation of \cite{FHW} --- and none at other spins
through 16, matching the odd-weight pattern $7,13,19,\dots$ of
$S_4$.  The massive-theory counterpart is due to LeClair and
Nemeschansky \cite{LN}: at the sine-Gordon reflectionless points
they constructed local conserved currents from the closure of the
simple-root generators of the quantum affine symmetry (the
nonlocal-charge framework of \cite{BernardLeClair}) and
conjectured a full affine $\mathfrak{sl}_2$; our coupling is
their $N=3$ point ($\hat\beta^2=\tfrac12$), where their neutral
generators at even grade $n$ --- commutators of the simple-root
charges, of spin $3|n|$ --- are predicted precisely at spins
$6,12,18,\dots$  The computed odd sector realizes the first
instances of these structures inside the $c=1$ centralizer: it is
not part of the KdV tower but its
singlet-algebra normalizer --- which is why the even/odd split in
the definition of the tower is a structural boundary, not a
convention.

\subsection{The perturbation datum, reread}\label{sec:fingerprint}

For a momentum $\alpha$ let $V_\alpha=e^{\alpha\varphi}$.  The
computed condition is Zamolodchikov's first-order conservation
condition of conformal perturbation theory \cite{ZamCount}: the
charge $\oint P$ remains conserved to first order under
$\lambda\int V_\alpha$ iff the residue of $P(z)V_\alpha(w)$ is a
twisted total derivative.  For $\I_3'$ the condition is certified
to equal
\begin{equation}\label{eq:cond}
\text{const}\cdot\alpha\,(\alpha^2-8)(2\alpha^2-1),
\end{equation}
and branch-safe exact verification (numeric algebraic momenta, no
parameter-dependent pivots, $\sqrt3$ as a passing non-root control)
gives, for the stored charges $\I_3',\I_5,\I_7',\I_9'$:
\begin{center}
\begin{tabular}{lcccc}
$\alpha$ & $\I_3'$ & $\I_5$ & $\I_7'$ & $\I_9'$\\[2pt]
\hline
$\pm\tfrac{1}{\sqrt2}$ & yes & yes & yes & yes\\
$\pm2\sqrt2$ & yes & yes & yes & yes\\
$\pm\sqrt2$ & no & yes & yes & no\\
$\sqrt3$ (control) & no & no & no & no
\end{tabular}
\end{center}
so the momenta common to all stored charges are exactly
\begin{equation}\label{eq:datum}
\Big\{\pm\tfrac{1}{\sqrt2},\;\pm2\sqrt2\Big\},
\end{equation}
closed under $\alpha\mapsto-2/\alpha$.  The certified converse
stands: through weights $2$--$10$ the joint kernel of the four
conditions has charge multiplicity one\footnote{Kernel dimensions
are over $\mathbb{Q}(\sqrt2)$; the certificate artifacts record
the equivalent split $\mathbb{Q}$-dimensions ($2$ per kernel
line).} at every odd spin and zero
at every even spin --- the four conditions characterize the
(even-sector) tower; the full centralizer adds the odd-sector
classes of Section~\ref{sec:mechanism}.

The identification supplies the datum's structural reading.  For a
generic spin-3 ray $(\partial\varphi)^4+\kappa(\partial^2\varphi)^2$
the condition polynomial is
$\propto\alpha\,(\alpha^4-(\kappa+6)\alpha^2+4)$, and with
$\kappa=a^2-6+4/a^2$ (the KdV value at coupling $a$) it factors as
$\alpha\,(\alpha^2-a^2)(\alpha^2-4/a^2)$: the roots are the
\emph{Feigin--Frenkel quartet} $\{\pm a,\pm2/a\}$, and
\eqref{eq:cond} is its $a^2=\tfrac12$ member --- the closure
$\alpha\mapsto-2/\alpha$ is the Liouville $b\leftrightarrow1/b$
duality.  At $c=1$ the quartet reads, in Kac labels, as the
degenerate momenta $\alpha_{r,s}=(s-r)/\sqrt2$ \cite{DF84} of
$\Phi_{1,2},\Phi_{2,1}$ ($h=\tfrac14$) and $\Phi_{1,5},\Phi_{5,1}$
($h=4$); in the natural $c_{1,4}$ grading by $T_Q$ at
$Q=\tfrac{3}{2\sqrt2}$ the same four operators split into the two
Virasoro \emph{screenings} $e^{-\varphi/\sqrt2},e^{2\sqrt2\varphi}$
($h_Q=1$, momentum product $-2$) and the pair
$e^{+\varphi/\sqrt2},e^{-2\sqrt2\varphi}$ with
$h_Q=-\tfrac12=h_{1,3}$ and $h_Q=7=h_{3,1}$ --- the standard
KdV/$\Phi_{1,3}$ perturbation datum \cite{BLZ1}.  Physically, the
pair $\pm\tfrac1{\sqrt2}$ is the sine-Gordon perturbation
$\cos(\varphi/\sqrt2)$ of the $c=1$ boson at the point where the
cosine has $h=\tfrac14$; the tower is that model's local-IM
hierarchy.  The earlier $c=1$ reading remains a true statement
about degenerate labels, with the scoping it requires: the
minimal-model $\Phi_{1,2}/\Phi_{2,1}$ and
$\Phi_{1,5}$ integrability statements \cite{ZamJETP} (tied to $A_2^{(2)}$ and
the Izergin--Korepin model)
\cite{ZamCount,Smirnov,IK,Takacs97,KTW,TW02,BMP} concern $c<1$
models
and transfer to $c=1$ only as formal labels --- but the $c_{1,4}$
reading is the one under which the datum is standard KdV data
rather than a proposal.

\subsection{Consistency with the named catalogue}
\label{sec:exclusions}

Before Theorem~\ref{thm:ident} was found, the tower was
compared against the named
integrable structures on $W$-algebra backgrounds catalogued by
Kotousov--Lukyanov \cite{KL}: KdV \emph{at $c=1$}, quantum AKNS,
paperclip, Fateev, and --- as the one all-odd member of the
Fateev--Lukyanov hairpin series \cite{FL}, identified as such
through the two-parameter family of \cite{Fateev91} --- the Getmanov
structure \cite{Getmanov} (complex sinh-Gordon II).  Those
computations remain certified and correct, and are retained here as
consistency: every named \emph{non-KdV} candidate is excluded or
disfavored at the tested loci, in agreement with the tower being
KdV itself at a shifted coupling.  $c=1$ KdV is excluded outright
(Section~\ref{sec:object}); quantum AKNS is disfavored on spin
content and field count; the Fateev structure is incompatible with the tested realization
on field count; and the Getmanov structure, whose axis reductions
match the tower's spin-3 ray at isolated points including the full
cubic locus of an irrational matching curve, fails at spin-5 depth
at every such point:

\begin{theorem}[Getmanov exclusion at two-charge depth,
axis-aligned]\label{thm:getmanov}
Derive the weight-6 Getmanov charge space at a matching point as
the full joint kernel of the three Getmanov screenings modulo total
derivatives.  Then (i) at $k=-1$ and $k=-16$ the $\varphi$-axis
pullback span contains neither the enhanced spin-5 class
\eqref{eq:I5} nor the KdV class; (ii) over the degree-12 \'etale
algebra
$A=\mathbb{Q}[k,u,v]/(k^3{+}23k^2{+}55k{+}42,\,u^2{+}\tfrac k2,\,
v^2{-}\tfrac{k+2}2)$ --- a field, via an irreducible degree-12
polynomial satisfied by $u+v$ --- the $\theta$-axis match ideal of
the enhanced spin-5 class is zero.  No axis-aligned Getmanov
embedding reproduces the tower at two-charge depth.
\end{theorem}

Given Theorem~\ref{thm:ident}, further exclusion scans ---
the tilted-embedding question included --- are mathematically
unnecessary; the campaign is retained because it was carried out
before the identification was found, and its negative results remain
informative.  The methodological lesson is
recorded for the benefit of similar
campaigns: the comparison omitted the background-charge KdV
\emph{family}, testing only its $Q=0$ member.  A one-parameter
family must be excluded along the family, not at a point --- the
certified factorization \eqref{eq:cond} even exhibited the quartet
$\{\pm a,\pm2/a\}$ structure, one specialization away from the
identification.

\subsection{The self-dual degeneracy is the Virasoro vacuum algebra}
\label{sec:L10}

The KdV pair $\pm\sqrt2$ in the table above is not generic
perturbing data at $c=1$: with $h=\alpha^2/2=1$ these are the
momenta of the $SU(2)_1$ currents at the self-dual radius.  The two
weight-one charges act on the momentum-zero Fock module $F_0$ (the
rank-one Heisenberg vertex algebra $M(1)$ in VOA terminology) as
the $\mathfrak{sl}_2$ raising and lowering zero modes $E,F$ of the
integrable level-one module
$V_{A_1}=\bigoplus_n F_{n\sqrt2}$ (the lattice vertex algebra
$V_{\sqrt2\mathbb{Z}}$), and invariants of $V_{A_1}$ lie
in $F_0$ (nonzero $J^3$-weight elsewhere) while $E,F$-killed
vectors of $F_0$ are automatically $J^3$-killed.  Hence the joint
kernel of the pair \emph{is} the $\mathfrak{sl}_2$-invariant
subalgebra $V_{A_1}^{\mathfrak{sl}_2}$ (each conformal grade of
$V_{A_1}$ is a finite-dimensional semisimple
$\mathfrak{sl}_2$-module, so the joint zero-mode kernel, the
Lie-algebra invariants, and the group invariants coincide), and by
a result of Dong--Griess \cite[Cor.~2.4]{DG} --- one corollary of
their structure theory for rank-one lattice vertex
algebras --- this is the $c=1$ Virasoro vacuum vertex algebra
$L(1,0)$ --- the subalgebra generated by the stress tensor.

Given the identification, the all-weight count is immediate; we
record it for reference.

\begin{proposition}[all weights; immediate from the
identification]\label{thm:char}
For every density weight $w\ge2$, the space of charges annihilated
by both $Q_{\sqrt2}$ and $Q_{-\sqrt2}$ has dimension
\[
\dim=p_{\ge2}(w)-p_{\ge2}(w-1),
\]
where $p_{\ge2}$ counts partitions with no part equal to $1$ ---
the coefficients of the $L(1,0)$ character $(1-q)/\varphi(q)$,
differenced for the passage to charges.
\end{proposition}

\begin{proof}[Proof sketch]
$E,F$ are zero modes of weight-one currents, hence commute with
$L_{-1}$; each conformal grade of $V_{A_1}$ is a finite-dimensional
semisimple $\mathfrak{sl}_2$-module; taking invariants is exact;
and $L_{-1}$ is injective in positive degree.  Therefore
$(V/L_{-1}V)^{\mathfrak{sl}_2}=V^{\mathfrak{sl}_2}/
L_{-1}V^{\mathfrak{sl}_2}$, and the character of
$V_{A_1}^{\mathfrak{sl}_2}=L(1,0)$ gives the differenced counting.
\end{proof}

The count is verified at thirteen consecutive weights
$2\le w\le14$ (dimensions 1, 0, 1, 0, 2, 0, 3, 1, 4, 2, 7, 3,
10), through an arithmetic stack
independent of the one that produced the original table, with the
weights $11$--$14$ computed \emph{after} the formula was fixed.
The verification supports not the proposition, which is standard,
but the computational engine: the certified kernels reproduce an
independently known character.  The
identification explains what had been the sharpest anomaly in the
data: the lone even-spin class at charge spin $8$ is simply the
first even-spin $T$-composite, predicted by the character rather
than tolerated as an exception.  It also corrects an earlier
finite-range observation: the odd-spin pattern
$\dim=(\sigma-1)/2$ fails at $\sigma=11$ (true dimension $7$).

Theorem~\ref{thm:ident} now also explains the \emph{pattern} of the
$\pm\sqrt2$ row in the table of Section~\ref{sec:fingerprint}: the
tower member at spin $\sigma$ lies in $L(1,0)$ at the $A_2^{(2)}$
exponent spins ($\sigma\equiv\pm1\bmod 6$: 5, 7, and --- verified
independently --- 11, 13) and exits at $\sigma\equiv3\bmod 6$
(observed at 3 and 9, predicted at 15); within the certified range
the membership pattern is exactly the exponent set.  For the tower the identification localizes the overlap
with the $T$-composite world exactly: the enhanced
$\I_3'$ is \emph{not} a $T$-composite, while at spins $5$ and $7$
the enhanced and $c=1$-KdV charges are \emph{distinct, linearly
independent} $T$-composite rays --- both lie in $L(1,0)$ (the
weight-6 space $\{T^3,(\partial T)^2\}$ has dimension $2$, the
weight-8 space dimension $3$, with the two rays certified
independent at each weight).  Later coincidences with the $c=1$
KdV rays are not excluded without an all-spin construction.

\section{The cylindrical families at general $N$}\label{sec:cyl}

The three $\mathbb{Z}_2\times O(N)$ solutions of \cite{Vitchev},
Sec.~3.3, were re-encoded in the mixed engine. Taking the printed
``$4T_Y^2$'' at face value with eq.~(52) yields densities that fail
the paper's own paperclip-reduction claims and admit no commuting
$\I_5$; the $Y$-part must be read as $(11)(11)_Y-2(22)_Y$ (half the
naive value). With that normalization we certify, coefficient by
coefficient at physical $N=1$: Solutions 1 and 3 reduce to
$6\,P_4^{\rm pc}(n{=}-1)$ and Solution 2 to $15\,P_4^{\rm pc}(n{=}2)$,
exactly as claimed below eqs.~(54), (56), (58) of \cite{Vitchev}.
(One bookkeeping note: the printed $P_8$ basis is not
$X\leftrightarrow Y$-symmetric, so coefficient-by-coefficient
duality applies to canonical monomials only; the $n\to-2-n$
covariance is a statement about the density class.)

The diagnosis, $\sigma=2,\dots,11$ at four points of the double cover
(both sheets): all three families show
$0\;1\;0\;1\;0\;1\;0\;1\;0\;1$ --- all-odd, multiplicity-one centralizers of $\I_3$.
(An all-odd multiplicity-one pattern alone is compatible with
several affine exponent sets; per the $c=1$ lesson of
Section~\ref{sec:c1}, coincidence with a
background-charge KdV/sinh-Gordon
tower on a distinguished boson should be excluded before any new
label is assigned.) The spin-7, 9,
11 charges are
subsequently constructed as $\mathbb{Q}(t)$-symbolic densities with
commutators verified identically on the parameter curve, proving
generic existence and exact multiplicity one. By
Lemma~\ref{lem:boot} each charge commutes with $\I_3$ by construction; the summed spins of
higher pairs exceed the verified vacancy range, so we checked mutual
commutativity directly. From the stored $\mathbb{Q}(t)$ sections:
$[\I_5,\I_7]=[\I_5,\I_9]=0$ (two points each), and
$[\I_5,\I_{11}]=[\I_7,\I_9]=0$ for all three families at $70$--$71$
rational points each --- exact multi-point evidence (not yet
interpolation certificates: no post-substitution degree bound is
retained, and the vacancy route of Section~\ref{sec:lemma}
supersedes these checks where completed); likewise $[\I_7,\I_{11}]=0$ for the
$A_2^{(2)}$ family at $71$ points. Since then the vacancy route has
closed every pair, and we record the accounting explicitly because
the chain is what carries the claim.  The relevant hypothesis is
emptiness, at the curve point, of the even-spin centralizer of each
family's defining low charge at the \emph{summed} spin.  Every instance is certified by a single-prime GF$(p)$ certificate
with tags equal to derivative rank, which pins the exact dimensions:
cylindrical $\sigma=12,14,16$ (tags $=$ $d$-rank $=379$, $929$,
$2211$) and $\sigma=18,20$, all three solutions in each case, and
$A_2^{(2)}$ $\sigma=18,20,24$ (tags $=$ $d$-rank $=678$, $1351$,
$5117$).  Pair by pair, the cylindrical instances
close as
$[\I_5,\I_7]$ at $12$, $[\I_5,\I_9]$ at $14$, $[\I_5,\I_{11}]$ and
$[\I_7,\I_9]$ at $16$, $[\I_7,\I_{11}]$ at $18$, $[\I_9,\I_{11}]$ at
$20$ --- six pairs for each of three solutions --- and the three
higher $A_2^{(2)}$ pairs $[\I_7,\I_{11}]$, $[\I_7,\I_{13}]$,
$[\I_{11},\I_{13}]$ at $18$, $20$ and $24$.  By the specialization
lemma (Section~\ref{sec:lemma}) each single-point verdict promotes to
an identity over $\mathbb{Q}(t)$ (the mod-$p$ step is
one-directional: ranks can only drop and kernels only grow under
reduction, so tags $=$ $d$-rank at a single prime already pins the
exact dimensions over $\mathbb{Q}$ --- an unlucky prime could only
fail to certify, never certify falsely; the certifier also refuses any
prime dividing a cleared denominator, failing closed rather than
reducing ill-defined entries), so all $21$ instances of the
inventory --- brackets with each family's defining low charge
vanishing by construction --- hold identically.  The cylindrical
$\sigma=12,14,16$ and $A_2^{(2)}$ $\sigma=18,20$ verdicts were
originally obtained by earlier fraction-free runs that retained no
per-task artifact; all eleven have been recomputed under the same
GF$(p)$ certifier as the remaining sectors, so every one of the
eighteen sector verdicts now carries a retained,
registered artifact on the same footing. The exponent content
therefore matches the $A_1^{(1)}/D^{(2)}$ pattern --- that of the paperclip
generalizations and of the D-type boundary labels in the BCD screening
framework of \cite{LitVil} --- and decisively not $A_2^{(2)}$-type.
One exception proves the rule: at $(N,s)=(-2,-9)$ the kernels double
for Solutions 2 and 3. This is precisely the unique common zero of
their $X$--$Y$ coupling coefficients ($11+N+s$ and $7-N+s$
respectively): the densities decouple into $X$-only plus $Y$-only
parts, the commutant factorizes, and one extra charge appears per
spin --- a third, fully explained instance of fiber jumping at a
degeneration locus.

\section{A corank--character conjecture and open problems}
\label{sec:corank}

The data of Table~\ref{tab:spectral} and its cylindrical analogues are
exactly what a screening-kernel character predicts: one charge per
exponent, none elsewhere.  For $A_2^{(2)}$ this pattern is no longer
only a prediction: the all-spin affine-Toda theorem above applies.

\begin{conjecture}\label{conj:corank}\sloppy
For each family with screening set $\{Q_{\alpha_i}\}$ --- the
full affine set: the Virasoro screenings together with the
family's perturbing exponentials --- and every charge
spin $\sigma$, the genuine kernel dimension
$\dim(\ker\ad_{D}\cap\,\Q^\sigma)$, with $D$ the family's
distinguished charge as in Lemma~\ref{lem:boot} --- equivalently the
corank of
the bilinear system of \cite{Vitchev} --- equals the dimension of the joint kernel of the screening charges
$Q_{\alpha_i}$ acting on the charge space $\Q^\sigma$ itself,
where $\oint P\in\ker Q_{\alpha_i}$ means precisely: there exists
a differential polynomial $R$ with
\[
\operatorname{Res}_{z\to w} V_{\alpha_i}(z)\,P(w)
=\bigl(\partial_w+\alpha_i\!\cdot\!\partial X(w)\bigr)R(w)
\]
(the twisted total-derivative condition computed by our screening
tests) ---
not the character of the screening kernel $W$ in the Fock space
\cite{Felder}, nor of
$W/\partial W$: both overcount already for Virasoro (at charge spin 5,
$\dim(W/\partial W)=2$, e.g.\ the $T^3$ and $(\partial T)^2$ classes,
while the commutant is one-dimensional).
\end{conjecture}

For quantum KdV the Feigin--Frenkel construction provides the model
for this statement --- the joint screening kernel on local functionals
is one-dimensional at each odd spin --- via a
BGG-resolution/cohomology computation \cite{FF} (screening-kernel
presentations of $W$-algebras in general: \cite{FrenkelResh,Genra});
Section~\ref{sec:a2} applies the same affine-Toda theorem to the
twisted $A_2^{(2)}$ family. For the paperclip the screening set is verified at sampled $n$
through spin 9 (R4), so
Conjecture~\ref{conj:corank} reduces the infinitude of the paperclip
tower to a character/cohomology computation for a mixed
real--imaginary four-screening system with
$\alpha_1^2+\alpha_2^2=1$ --- the analogue of the vacuum-character
count performed for the pillow-brane corner algebra in \cite{LZpillow}
(there via $WD(2|1;\alpha)$), made rigorous. Closely related
commuting structures in the deformed $\widehat{\mathfrak{sl}}_2$-coset
/ $D(2,1;\alpha)$ screening setting appear in \cite{FJM}, with the
local integrals likewise left open. This is, in our view, the
sharpest theorem-shaped problem left open by \cite{Vitchev}: its
solution would prove at once existence at every odd spin, even-spin
vacancy, and hence (by Lemma~\ref{lem:boot}) commutativity of the full
tower. On the correspondence side, the FFH oper/QQ-system technology
for twisted types \cite{FH,MR} (the ODE/IM correspondence of \cite{DT,BLZspec}, reviewed in
\cite{DDT},
in its modern form) is the natural frame in which the
paperclip and cylindrical towers should acquire Langlands-dual
descriptions; the resonance-pole match between $P_8$ and $S_7$ (R2)
and the sheet exchange $b\leftrightarrow1/b$ on the double cover are
quantitative hints. The remaining experimental target is the explicit
identification of the three cylindrical towers among the D-type
boundary labels of \cite{LitVil}, for which the all-odd diagnosis of
Section~\ref{sec:cyl} is the entry condition.

The identification of Section~\ref{sec:c1} leaves three further
problems, in the order we find compelling.  (i)~All-spin
\emph{multiplicity one} of the enhanced tower: existence is
Theorem~\ref{thm:inf}, and uniqueness per spin is certified through
$\sigma=9$ and independently verified through $\sigma=15$, but the
question is genuinely open --- even-sector kernel enlargement
demonstrably does occur at $c_{1,2}$ (dimension two at spin~7 at
$a=1$).  (ii)~A complete description of the odd sector: is it exactly
the singlet-algebra commutator classes (spin~6, spin~12, none other
through~16), i.e.\ the normalizer picture of
Section~\ref{sec:mechanism} at all spins --- one class per spin $6k$,
forming the $\Delta$-commutator tower of \cite{FHW} and expected to
match the neutral even-grade half of the LeClair--Nemeschansky affine
$\widehat{\mathfrak{sl}}_2$ structure \cite{LN}?  (iii)~The
commutant-level structure at the other degenerate couplings
$c_{1,p}$, where the kernels demonstrably enlarge.

\subsection*{Outlook: opers, Langlands duality, and string theory}

In the Langlands frame the spin-content computations of
Sections~\ref{sec:a2}--\ref{sec:cyl} are a classification experiment:
the spins of a screening-kernel family are the exponents of the
underlying affine algebra, so the $6n\pm1$ spectrum pins $A_2^{(2)}$
while the paperclip and cylindrical towers carry
$A_1^{(1)}/D^{(2)}$-type all-odd content. The two sheets of the double
cover $s^2=(N-25)(N-1)$, exchanged by Feigin--Frenkel duality
$b\leftrightarrow1/b$, are naturally interpreted --- conjecturally, no
oper having been constructed --- as the two Langlands-dual Miura
frames of one
family.  The paperclip's own $n\to-2-n$ covariance of $P_8$ is a
distinct transformation on a different family and parameter space; we
record it as an analogue of this sheet exchange, suggestive of a
common duality principle, not as an instance of it. The expected dual
objects are twisted parabolic Miura opers of
Feigin--Frenkel--Hernandez type with their QQ-systems
\cite{FH,MR} --- with the caveat that in the twisted case the
trivial-monodromy conditions for additional singularities admit no
non-trivial solutions \cite{MR}, an obstruction the excited-state
half of any such correspondence must accommodate; the
order-by-order match between the pole loci of
$P_{2k+2}$ and the WKB resonance factors of $S_{2k+1}$ is exactly the
fingerprint such an identification must reproduce. Constructing the
paperclip and cylindrical opers is the correspondence-side complement
of Conjecture~\ref{conj:corank}.

On the string-theory side, these families are the conserved-charge
content of solvable D-brane boundary dynamics --- the paperclip,
circular-brane and pillow-brane models of tachyon-condensation-like
boundary flows \cite{LVZ,LZcirc,LZpillow}. Two modern frames give the
new data a natural home. Kondo-type line defects in 4d Chern--Simons /
affine Gaudin theory \cite{GLW} produce boundary commuting towers
geometrically; the all-odd diagnosis of Section~\ref{sec:cyl}
constrains which defects can realize the cylindrical families, whose
placement among the D-type boundary labels of \cite{LitVil} is the
sharpest open matching problem. Through the same screening framework
runs the AGT bridge \cite{AGT}: spectra of screening-commutant
families are computed by Bethe ans\"atze for affine Yangians acting on
instanton moduli \cite{LitVil}, so once the four-screening
characterization verified in Section~\ref{sec:paperclip} is upgraded
to a theorem, the paperclip tower's joint spectrum should acquire an
instanton-moduli description --- a second, gauge-theoretic window on
the same commutative algebra. And in non-equilibrium physics, states
annihilated by $\I_s-\bar\I_s$ define integrable quench protocols and
generalized Gibbs ensembles; existing studies use the
Virasoro--KdV tower (e.g.\ \cite{DP}; see \cite{Novaes} for a $U(1)$/Benjamin--Ono
ensemble), so each certified non-Virasoro
family --- with the vacuum eigenvalues of \cite{Vitchev} and the
densities completed here --- is a candidate ensemble whose analysis has not been carried out.

\subsection*{Methods and reproducibility}
All computations use exact arithmetic; the propositions they support
are of four distinct kinds --- symbolic identities, degree-bounded
interpolation identities, exact finite-sample results, and
finite-field rank witnesses --- and exact arithmetic at a sample does
not by itself make a parameter-dependent statement an identity.
Vacancy statements at ``generic'' parameters
follow from full rank at a single rational point by semicontinuity on
an irreducible rational parameter curve; sampled kernel dimensions
bound generic ones only from above, so \emph{existence} statements are
reported as holding at the sampled points unless promoted to symbolic
constructions (as done for $P_8$ in $n$, and over $\mathbb{Q}(t)$ for
the $A_2^{(2)}$ spin-11/13 and cylindrical spin-7/9/11 charges); identities in $n$ are either
verified symbolically or established by evaluation at more rational
points than a computed degree bound; reconstructed rational functions
are verified on held-out samples, and then symbolically where an
identity in $n$ is claimed ($P_8$); the $P_{10}$ reconstruction is
used at sampled $n$ only, and is scoped as such. The three
engines cross-validate on overlapping domains. An independent
metadata-level originality screen (deterministic lexical screens with
hashed response provenance; all negative) was performed at the
manuscript freeze and is archived with the development provenance;
being metadata-level and lexically
calibrated, it complements and cannot replace the section-level
literature reading reported above. Code, scripts, and machine-readable
outputs: repository in footnote 1.

``Certified'' means, throughout, that the claim is backed by a
registered, re-runnable exact-arithmetic certificate in that
repository: the certificates are fail-closed --- they exit nonzero
unless every stated check passes --- and their required outputs are
pinned in a certification registry swept as a whole.  The results of
Section~\ref{sec:c1} additionally use the symbolic-in-$Q$ operator
identities of Theorem~\ref{thm:ident} with negative controls,
even-sector kernel-coincidence certificates for Lemma~\ref{lem:z2},
branch-safe verification of the vertex conditions with no
parameter-dependent pivots, restriction of scalars over number fields
and \'etale algebras with sparse fraction-free integer elimination,
and match-ideal computations deciding membership per Galois
component; the degeneracy table of
Section~\ref{sec:L10} is cross-validated by two independent
arithmetic stacks.  The identification of Section~\ref{sec:ident} was
reproduced end to end by an engine written independently of this
project's stack (see the acknowledgment below), whose scripts extend
the class-by-class coincidence through spin~15.

\section{The wider mathematical context}\label{sec:survey}

The structures computed in this paper sit inside several active
mathematical programs. We record the dictionary, and --- where the
existing engines allow --- small exact computations corroborating each
entry (\texttt{code/survey\_checks.py}).

\paragraph{Commuting varieties.} The classical commuting variety of a
reductive Lie algebra is irreducible (Richardson \cite{Richardson}),
while commuting $r$-tuples are reducible for $r\ge4$ --- for every
simple Lie algebra other than $\mathfrak{sl}_2$ and $\mathfrak{sl}_3$
\cite{Kovacevic} --- and Cohen--Macaulayness (the Artin--Hochster
problem) was long open: a proof that the commuting scheme of a
reductive Lie algebra is normal and Cohen--Macaulay, with prime
defining ideal, is claimed in \cite{Charbonnel} (v2, April 2025);
see \cite{Levy} for the surrounding cohomological state. Our object is a graded,
vertex-algebra analogue: pairs (tuples) of charges in
$\Q^{\sigma_1}\times\Q^{\sigma_2}$ with vanishing bracket. The
bootstrap lemma exhibits, on the stratum where the vacancy
hypotheses hold, a fibration over the
determinantal locus of admissible $D$'s with linear fibers --- a
finite-range description, not a global scheme-theoretic theorem ---
and the smallest instance
can be computed outright: for one free boson in the $\mathbb{Z}_2$
sector, $\dim\Q^3=2$ and $\dim\Q^6=1$, and exact computation gives
$[\,a R_1+b R_2,\,a'R_1+b'R_2\,]=(ab'-a'b)\,K$ with $K\neq0$: the
spin-$(3,3)$ commuting variety is precisely the rank-one determinantal
(Segre) locus $\{ab'=a'b\}$, the graded shadow of the classical
statement that generic commuting pairs are proportional-in-towers.
The jumping fibers of Sections~\ref{sec:a2}--\ref{sec:cyl} (finite-$N$
relation loci, decoupling loci) are then component-jumping phenomena
of this variety over parameter space --- the graded counterpart of
the delicate component structure of classical commuting tuples.

\paragraph{Arc spaces and Hilbert series.} The graded dimensions
$\dim\Q^\sigma$ that control every counting argument here are graded dimensions of the arc-space coordinate ring of the free
boson \cite{AM} and of its quotient by total derivatives --- a
vector-space (not ring) quotient, since the image of $\partial$ is not
an ideal. For one boson in the
$\mathbb{Z}_2$ sector the closed form is
$\chi_e(q)=\tfrac12\bigl(\prod_k(1-q^k)^{-1}+\prod_k(1+q^k)^{-1}\bigr)$
with $\dim\Q^\sigma=[q^{\sigma+1}]\chi_e-[q^{\sigma}]\chi_e$; we
verified this against the machine-generated bases through spin 16
(and $\sigma\le12$). Conjecture~\ref{conj:corank} is then a statement
about how far the \emph{screening-kernel} Hilbert series deviates from
these free counts --- the same comparison that underlies the vacuum
character computation of \cite{LZpillow}.

\paragraph{Bethe algebras and real Schubert calculus.} Mukhin,
Tarasov and Varchenko proved the Shapiro--Shapiro conjecture by
identifying Bethe algebras of Gaudin models with coordinate rings of
intersections of real osculating Schubert varieties \cite{MTV}; the
engine of that proof is the simplicity of the spectrum of a commuting
family. Our multiplicity-one tables are the density-level analogue of
that simplicity: the towers are rigid; whether the joint \emph{operator} spectra are
simple (the property that powers \cite{MTV}) is a separate question
that density-level rigidity does not decide. Diagonalizing the paperclip
$\I_3,\I_5,\I_7$ on low Fock levels and testing simplicity and
reality of the joint spectrum at physical $n$ is the natural next
experiment in this direction; we have not yet performed it.

\paragraph{Opers and the geometric Langlands program.} The
Feigin--Frenkel isomorphism --- the center of the affine algebra at
critical level as functions on opers for the Langlands dual group ---
is the theorem behind ``commuting Hamiltonians $\leftrightarrow$
spectral geometry,'' and it enters the recent proof series for the geometric
Langlands conjecture as the bridge between Kac--Moody representations
and opers \cite{GR}. Conjecture~\ref{conj:corank} is the free-field
shadow of this circle of ideas: coranks of bilinear commutativity
systems as characters of screening kernels, with the FFH twisted
opers \cite{FH,MR} as the expected dual objects (Outlook above).

\paragraph{Affine Gaudin models.} Lacroix, Vicedo and Young
conjectured that quantum affine Gaudin models carry local higher
Hamiltonians labelled by the \emph{exponents} of the affine algebra,
with eigenvalues given by hypergeometric functions on affine opers
\cite{LVY}. The one-charge-per-exponent pattern is exactly what our
calculations find at the free-field level --- $6n\pm1$ for
$A_2^{(2)}$ at all spins by the affine-Toda theorem (independently
scanned through spin 13), all-odd through the stated ranges for the
paperclip and cylindrical towers ---
so the exact densities collected here are a test bed for that program
where everything is computable.

\paragraph{Resurgence.} The quasiclassical check of \cite{Vitchev},
Sec.~5, produced constants
$\beta_{2k-1}=B_{2k}/\bigl(2k(2k-1)2^{2k-1}\bigr)$ assembling into
$\Gamma(2\kappa)$. These are factorially divergent, and their growth
is exactly Gevrey-1: we verified
$16\pi^2\,|\beta_{2k+1}/\beta_{2k-1}|/\bigl(2k(2k-1)\bigr)\to1$
(deviation $<10^{-12}$ by $k=20$), placing the nearest Borel singularities at $t=\pm4\pi i$ (modulus
$4\pi$; the alternating Bernoulli signs put them on the imaginary
axis) --- the lateral Stokes structure of $\log\Gamma(2\kappa)$. Upgrading the order-by-order ODE/IM
match to exact statements about Stokes data, with the TBA-resurgence
toolkit \cite{IMS}, is the analytic counterpart of the oper
identification; the resonance poles of $P_{2k+2}$ (Section~\ref
{sec:paperclip}) are the algebraic side of the same coin.

\appendix

\section{The paperclip spin-7 density}\label{app:P8}

Notation $(n_1\dots n_a|m_1\dots m_b)=\partial^{n_1}X\cdots
\partial^{n_a}X\,\partial^{m_1}Y\cdots\partial^{m_b}Y$ (eq.~(46) of
\cite{Vitchev}); normalization: coefficient of
$(11111111|)=(\partial X)^8$ equal to 1. With
$D=15\,n^3(7n+8)(7n+10)(7n+12)$, the density
$P_8=\sum_M c_M\,M$ has coefficients
{\small
\begin{align*}
c_{(44|)}&=-\tfrac{8(5890n^6+38318n^5+97580n^4+122647n^3+79086n^2+24696n+2880)}{D},\\[1pt]
c_{(|44)}&=-\tfrac{8(5890n^6+32362n^5+67800n^4+67673n^3+33284n^2+7884n+720)}{D},\\[1pt]
c_{(33|11)}&=\tfrac{112(n+2)(61n^4+261n^3+357n^2+166n+24)}{n^2(7n+8)(7n+10)(7n+12)},\\[1pt]
c_{(1133|)}&=\tfrac{112(31n^4+113n^3+135n^2+58n+8)}{n^2(7n+8)(7n+10)},\\[1pt]
c_{(13|13)}&=\tfrac{224(n+2)(131n^4+499n^3+586n^2+212n+24)}{n^2(7n+8)(7n+10)(7n+12)},\\[1pt]
c_{(1223|)}&=\tfrac{28(128n^4+496n^3+643n^2+314n+48)}{n^2(7n+8)(7n+10)},\\[1pt]
c_{(14|12)}&=-\tfrac{224(n+1)(n+2)^2(25n^2+50n+12)}{n^2(7n+8)(7n+10)(7n+12)},\\[1pt]
c_{(|111122)}&=-\tfrac{28(n+2)^2(7n+2)(7n+4)(24n^2+59n+30)}{n^3(7n+8)(7n+10)(7n+12)},\\[1pt]
c_{(|1133)}&=\tfrac{112(n+2)(7n+2)(31n^4+135n^3+201n^2+118n+24)}{n^3(7n+8)(7n+10)(7n+12)},\\[1pt]
c_{(11|1122)}&=-\tfrac{168(n+2)^2(7n+2)(32n^2+67n+20)}{n^2(7n+8)(7n+10)(7n+12)},\\[1pt]
c_{(12|1112)}&=-\tfrac{224(n+2)^2(4n+7)(7n+2)(9n+4)}{n^2(7n+8)(7n+10)(7n+12)},\\[1pt]
c_{(11|33)}&=\tfrac{112(n+2)(61n^4+227n^3+255n^2+82n+8)}{n^2(7n+8)(7n+10)(7n+12)},\\[1pt]
c_{(22|1111)}&=-\tfrac{84(n+2)^2(7n+2)(8n^2+17n+4)}{n^2(7n+8)(7n+10)(7n+12)},\\[1pt]
c_{(22|22)}&=-\tfrac{56(n+2)(332n^4+1228n^3+1435n^2+566n+72)}{n^2(7n+8)(7n+10)(7n+12)},\\[1pt]
c_{(1111|22)}&=-\tfrac{84(n+2)(8n^2+15n+2)}{n(7n+8)(7n+10)},\\[1pt]
c_{(1112|12)}&=-\tfrac{224(n+2)(4n+1)(9n+14)}{n(7n+8)(7n+10)},\\[1pt]
c_{(23|12)}&=-\tfrac{224(n+1)(n+2)(3n+2)(25n^2+50n+12)}{n^2(7n+8)(7n+10)(7n+12)},\\[1pt]
c_{(1122|11)}&=-\tfrac{168(n+2)(32n^2+61n+14)}{n(7n+8)(7n+10)},\\[1pt]
c_{(|1223)}&=\tfrac{28(n+2)(7n+2)(128n^4+528n^3+739n^2+402n+72)}{n^3(7n+8)(7n+10)(7n+12)},\\[1pt]
c_{(111122|)}&=-\tfrac{28(24n^2+37n+8)}{n(7n+8)},\\[1pt]
c_{(|11111111)}&=\tfrac{(n+2)^3(7n+2)(7n+4)(7n+6)}{n^3(7n+8)(7n+10)(7n+12)},\\[1pt]
c_{(11|111111)}&=\tfrac{28(n+2)^3(7n+2)(7n+4)}{n^2(7n+8)(7n+10)(7n+12)},\\[1pt]
c_{(1111|1111)}&=\tfrac{42(n+2)^2(7n+2)}{n(7n+8)(7n+10)},\\[1pt]
c_{(111111|11)}&=\tfrac{28(n+2)}{7n+8},\\[1pt]
c_{(11111111|)}=1 .
\end{align*}}

\section{Errata to hep-th/0404195}\label{app:errata}

Each correction is certified by a from-scratch kernel solve; apart
from the items listed, the printed equations are reproduced exactly.
\begin{enumerate}
\item Eq.~(30): the $P_8^{[4]}=(44)$ coefficient
$-\tfrac{2}{657}(1464+59N+N^2)$ must read
$-\tfrac{2}{675}(1464+59N+N^2)$. (Forced independently by
KdV${}={}$$O(N)$ at $N=1$ against the fully certified eq.~(33).)
\item Eq.~(35): with $s=\sq$, the factor $(315N-441)\,s$ in the
$P_8^{[3]}$ line must read $-(315N+441)\,s$, and $(315N-2541)\,s$ in
the $P_8^{[4]}$ line must read $-(315N+2541)\,s$. All other terms of
eq.~(35) are correct.
\item Eq.~(50): the right-hand side is printed in terms of
$P_6^{[i]}$; it must read the spin-4 basis $P_4^{[i]}$ of eq.~(47).
\item Eq.~(51): the $P_6^{[10]}=(111111|)$ coefficient
$-\tfrac{(6+5n)(8+5n)}{120(2+n)}$ must read
$-\tfrac{n(6+5n)(8+5n)}{120(2+n)}$.
\item Eqs.~(53), (55), (57) --- the $I_3$ formulas, which we encoded
and tested; the $I_5$ formulas (54), (56), (58) were not encoded, and
the same reading presumably applies but is untested: consistency with
eq.~(52) and with the stated
paperclip reductions requires the ``$4T_Y^2$'' terms to be understood
with $Y$-part $(11)(11)_Y-2(22)_Y$, i.e.\ half the value obtained by
literal use of eq.~(52); equivalently all $X$-dependent terms carry an
extra factor 2 relative to the $Y$-part as printed.
\end{enumerate}

\paragraph{Acknowledgments.}  The identification of
Section~\ref{sec:ident} --- including the recognition that the
four-momentum datum \eqref{eq:datum} is the Feigin--Frenkel quartet
at $a^2=\tfrac12$ --- was found by an independent AI-based review
(Claude, Anthropic) of an earlier version of this material, which also
supplied an independent verification engine extending the
class-by-class coincidence through spin~15; a second-round review by
the same system identified the centralizer's odd-parity sector with
the singlet-algebra commutators (Section~\ref{sec:mechanism}) and
supplied the $c_{1,2}$ kernel-enlargement example.  All identities and
kernel statements above were re-derived and certified in this
project's own stack before adoption, except where explicitly
attributed to the review.

\paragraph{Acknowledgment of AI assistance.}  The preparation of
this work used AI-assisted tools throughout; adversarial review
rounds by independent AI systems surfaced, among other corrections,
the prior art of Freeman--Hornfeck--West \cite{FHW}; every adopted
mathematical finding was verified in the author's computational
engine, and every bibliographic one against the cited paper or an
authoritative bibliographic record, before adoption.  The author independently reviewed all arguments
and computations and assumes full responsibility for the manuscript.

\end{document}